# Investigating the Significance of the Bellwether Effect to Improve Software Effort Prediction: Further Empirical Study

Solomon Mensah[1], Jacky Keung[1], Stephen G. MacDonell[2], Michael Franklin Bosu[3], and Kwabena Ebo Bennin[1]

[1] *Department of Computer Science, City University of Hong Kong, Hong Kong*
*(smensah2-c@my. cityu.edu.hk; jacky.keung@cityu.edu.hk; kebennin-c@my.cityu.edu.hk)*
[2] *Department of Information Science, University of Otago, Dunedin 9016, New Zealand, and Department of IT and Software Engineering, Auckland University of Technology, Auckland 1010, New Zealand*
*(stephen.macdonell@otago.ac.nz)*
[3] *Centre for Business, Information Technology and Enterprise, Wintec, Hamilton 3200, New Zealand*
*(michael.bosu@ wintec.ac.nz)*

**Abstract**

*Context:* In addressing how best to estimate how much effort is required to develop software, a recent study found that using exemplary and recently completed projects [forming Bellwether moving windows (BMW)] in software effort prediction (SEP) models leads to relatively improved accuracy. More studies need to be conducted to determine whether the BMW yields improved accuracy in general, since different sizing and aging parameters of the BMW are known to affect accuracy. *Objective:* To investigate the existence of exemplary projects (Bellwethers) with defined window size and age parameters, and whether their use in SEP improves prediction accuracy. *Method:* We empirically investigate the moving window assumption based on the theory that the prediction outcome of a future event depends on the outcomes of prior events. Sampling of Bellwethers was undertaken using three introduced Bellwether methods (SSPM, SysSam, and RandSam). The ergodic Markov chain was used to determine the stationarity of the Bell-wethers. *Results:* Empirical results show that 1) Bellwethers exist in SEP and 2) the BMW has an approximate size of 50 to 80 exemplary projects that should not be more than 2 years old relative to the new projects to be estimated. *Conclusion:* The study's results add further weight to the recommended use of Bellwethers for improved prediction accuracy in SEP.

## NOMENCLATURE

*A. Acronyms*

| | |
|---|---|
| DNN | Deep neural networks. |
| BIC | Bayesian information criteria. |
| BMW | Bellwether moving window. |
| EMC | Ergodic Markov chain. |
| GP | Growing portfolio. |
| MAE | Mean absolute error. |
| MCMC | Markov chain Monte Carlo. |
| RandSam | Random sampling. |
| SEP | Software effort prediction. |
| SSPM | Stratification, sampling, and prediction method. |
| SysSam | Systematic sampling. |
| TPM | Transition probability matrix. |

*B. Notation*

| | |
|---|---|
| $\delta$ | Cliff's delta effect size. |
| $\Omega$ | Sample space. |
| $f_{new}$ | Feature set of new projects. |
| $f_{bell}$ | Feature set of Bellwether projects. |
| $w_i$ | Projects within the *ith* window. |
| $P_{ij}^n$ | Probability that state *i* will move to *j* after *n* steps. |
| $D$ | Number of historical projects from a dataset. |
| $N$ | Number of projects sorted in ascending order. |
| $m$ | Subset of recently completed projects. |
| $q$ | Number of partitions or clusters. |
| $p_i$ | Hold-out project(s) whose target is to be estimated. |

## 1. INTRODUCTION

Sep is the process used in predicting the effort needed for software development, thus supporting scheduling, costing, and the allocation of resources to meet project delivery deadlines [1]–[3]. According to Menzies *et al.* [4], accurately predicting software effort is of importance since it minimizes underestimation and overestimation of software effort, both of which can result in serious problems for a company [5]. Irrespective of how complex or simple an estimation model is, prediction accuracy can be affected by the sampled data used for training and validating the model. Thus, there is a need to ensure the selection of relevant training and validation sets when building a robust and relatively accurate prediction model.

The existence and appropriate sampling of exemplary projects (considered as the training set and) referred to as *Bellwethers* has successfully been used to improve prediction accuracy in some aspects of software engineering. Krishna *et al.* [6] demonstrated the existence of *Bellwethers* in the context of software defect prediction by making an implicit assumption that, *given N cases of defect prediction projects, there exists exemplary project data that can yield a relatively high prediction accuracy on the remaining project cases*. They considered their approach of finding the *Bellwether* as a simple transfer learning method [7], whereby using exemplary projects yielded improved prediction accuracy for new projects to be estimated. Similarly, in previous studies by Chen *et al.* [8], [9], a subset of sales of an item was shown to provide a successful basis for the prediction of the item's annual sales. Chen *et al.* [8], [9] obtained *Bellwethers* to be used for predicting annual sales of a new item by first sampling existing weekly sales data based on a low data-collection cost. The existence of such exemplary observations from a given repository to be considered as a *Bellwether* for making predictions of new cases is referred to as the *Bellwether effect* [6] .

Based on the MCMC methodology [10], the outcome of an event in an experiment (in this case, a new project) depends on those of previous experiments (in this case, recently completed projects with defined transition probabilities). Using theoretical and empirical analysis, we argue that MCMC modeling based on a novel step-by-step method can be used to sample relevant and recently completed exemplary projects to achieve relatively higher prediction accuracy than would be the case otherwise. We refer to this subset of exemplary and recently completed projects as the BMW [11].

Such an approach is a specific instance of the recent focus in SEP on selecting the most relevant, reliable, and recent data samples from historical data repositories and their inclusion strategies for predictive modeling [12]–[15]. According to Lokan and Mendes [16], [17] , selection of a recently completed training set (a *moving window*) for model building from a set of chronologically arranged projects can improve prediction accuracy in SEP. The *moving window* theory is based on the assumption that recent projects are more likely to share similar characteristics with new projects. Recent studies [11], [12], [18] support the theory of Lokan and Mendes that the use of a *moving window* as the training set improves the prediction accuracy of SEP models. This *moving window* approach has recently been supplemented with weighting functions, which have been shown to *further improve the accuracy of SEP models*. This theory was postulated by Amasaki and Lokan [12], [19] and was empirically confirmed when using large-sized moving windows. Taken overall, these prior studies found SEP models built using recently completed project data to be superior to SEP models that used all available historical project data (commonly referred to as a *GP* ).

Irrespective of the significant improvements that can be achieved by using weighted moving windows, challenges still exist in determining how large and how old the window should be prior to predicting new project cases. The *window size* simply refers to the number of recently completed projects that is most appropriate for building the SEP model. The *window age* refers to the elapsed time or duration span of projects (forming the window) that has existed for not more than $t$ calendar years or months [13].

Previous studies [12], [15], [19] considered using window sizes of 20, 30, 40 … 120 projects when building prediction models. After assessing prediction accuracy, they selected the best window size to predict the software effort (the target variable) of the *new* project(s). Similarly, projects forming different window ages (such as 1 year, 1.5 years, 2 years, 2.5 years, …) were also considered by previous studies [7], [9] when building prediction models. These sizing and aging parameters can significantly affect prediction accuracy irrespective of the weighting functions used [12], [15]. This is mainly due to the variations in the sample sizes, project ages, and different chronological datasets used in modeling.

A chronological dataset is composed of projects from a given repository that have respective development start and completion dates. Only some publicly available repositories have these dates included—the most well-known examples related to SEP are the Kitchenham dataset [20], the International Software Benchmarking Standards Group (ISBSG) dataset [14], and the Finnish dataset [14]. Recent studies [15], [18] that used the ISBSG dataset found the use of moving windows to improve prediction accuracy when compared to using a GP. On the other hand, in the case of the Finnish dataset, the moving windows assumption proved to be reliable in a recent study by Amasaki and Lokan [12] but unreliable in a further study by Lokan and Mendes [14]. Thus, while most studies have indicated the superior accuracy of moving windows, there have been some exceptions. Even though there is a challenge to the use of moving windows based on this previous study's [14] findings, we maintain that effective sampling of a subset of exemplary projects should and can result in improved prediction accuracy, as demonstrated in our original study [11]. Thus, more studies with different datasets and different modeling techniques are needed to further investigate the utility of moving windows and its generalization in SEP. That is, to investigate *if the sampling of exemplary projects in a BMW with defined window size and age can be modeled based on Markov chains to further improve prediction accuracy*. This paper makes use of the concept of the *Bellwether effect* to sample the moving window for predicting the software effort of new project cases.

The objective of this study is to theoretically and empirically evaluate the utility of *Bellwethers* with defined size and age parameters in terms of leading to improved accuracy in SEP. We make use of chronological datasets from the ISBSG and the PROMISE [21] repositories in our empirical analysis. We address the window sizing and aging constraints by introducing three *Bellwether methods:*

1) an *SSPM*;
2) a *SysSam* method;
3) a *RandSam* method.

This study makes three contributions.

1) We theoretically and empirically demonstrate the existence of *Bellwethers* with defined window size and age for SEP.
2) We introduce three *Bellwether methods* for sampling *Bellwethers* for improved prediction accuracy.
3) We provide a guideline for sampling *Bellwethers* for its practical use in prediction modeling.

The remaining sections of the paper are organized as follows. Section 2 presents the background of the study. Section 3 presents the key concept of the *Bellwether*, moving windows, GP, Markov chain Monte Carlo method, central limit theorem, law of large numbers and four robust statistical measures utilized in this study. Section 4 presents the postulations with their respective proofs for sampling *Bellwethers*. The *Bellwether methods* are discussed in Section 5. Section 6 details the methodological procedure employed. Section 7 presents our experimental results and the discussion emanating from the empirical analysis. Section 8 provides a general guideline for incorporating *Bellwethers* for prediction purposes. Section 9 presents a summary of related works pertaining to moving windows. Section 10 presents the threats to the validity of our study, and Section 11 concludes this work and discusses future directions.

## 2. BACKGROUND OF STUDY

This paper represents a substantive extension of one of our prior research efforts. In this section, we describe the original study [11] and its results as well as the goals and distinctive elements of the current study, following the replication guidelines introduced by Carver [22].

**A. Original Study**
The main goal of our original study [11] was to investigate the existence of exemplary projects (i.e., the *Bellwether effect*) in SEP, focusing on chronological projects. We investigated whether the use of exemplary and recently completed projects in a BMW could improve prediction accuracy as compared to using the entire collection of historical projects over time (*GP*). Thus, given $p_i$ projects (the hold-out set) whose target values are to be estimated, we investigated whether an entire set of *N* historical projects or a subset of *m* recently completed projects ($m \subset N$) can yield improved prediction accuracy. Note that both the *GP* (*N*) and the BMW (*m*) had their development completion dates prior to the start dates of the $p_i$ projects.

The original study [11] used two chronological datasets, namely ISBSG dataset (release 10) and the Kitchenham dataset [20]. With regard to the ISBSG dataset, a set of 1097 pre-processed cross-company projects were used for the empirical analysis. A set of 142 preprocessed single company projects were considered from the Kitchenham dataset. We applied two data transformation techniques, namely *log transformation* and *z-score normalization*, to investigate whether they affected pre- diction accuracy of the estimates.

The original study was initialized with the formulation of six postulations that were theoretically proven and empirically validated with the two studied datasets. A novel *Bellwether method* [23] based on MCMC was used to sort, stratify, and sample relevant projects that constituted the BMW. Here, an initial baseline of recently completed projects was used as the training set whilst the remaining strata of projects were considered as the validation set. The *X-means* clustering algorithm was used to partition the sample space into *q* clusters. The baseline, $w_q$ was updated with projects from $w_{q-1}$ until $w_q^*$ yielded the best prediction accuracy based on three evaluation measures [MAE, mean balanced relative error (MBRE), and mean inverted balanced relative error (MIBRE)]. Predictions were generated based on three learners, namely the automatically transformed linear model (ATLM) [24], a DNN model [25], and ordinary least squares regression [26]. After obtaining the BMW with the best prediction accuracy, it was benchmarked against the *GP* to predict the target values of the $p_i$ hold-out projects. Note that the $p_i$ projects were not considered in the training and validation sets.

The main results obtained from the original study [11] are as follows.

1. The *Bellwether effect* was evident in both the ISBSG and Kitchenham datasets.
2. We found that *Bellwethers* existed in populations of size not less than 100 chronological projects from both datasets.
3. With regard to window age, the BMW sampled from the ISBSG dataset had a relative effective size of not more than 2.5 calendar years. Similarly, the age of the BMW from the Kitchenham dataset was not more than 2 calendar years.
4. With regard to window size, the number of exemplary projects forming the BMW from the ISBSG dataset was 257 (23.4%) and that of the Kitchenham dataset was 87 projects (61.3%).
5. In the case of the three learners, we noted that the DNN yielded the best prediction accuracy in both datasets. This was followed by the baseline SEP model (i.e., ATLM [24]).
6. On average, we found that the BMW yielded a relatively higher prediction accuracy as compared to the *GP* in both datasets.
7. We realized a significant prediction improvement when the Gaussian weighting function was applied on the BMW in both datasets.
8. The *log transformation* resulted in improved prediction accuracy as compared to the *z-score normalization* in both cases.

**B. Current Study**
In order to generalize the usage of BMW for improved SEP accuracy, there is a need to conduct further empirical studies that could justify the theoretical postulations in the original study [11], provided the results are consistent.

Alternatively, if the results are not consistent (i.e., the BMW yields no or a less significant effect in the current study), then insights can be gained in terms of the variations between the datasets investigated across the two studies. The specific intent of this study is to investigate the possible existence and effect of *Bellwethers* as well as the sizing and aging parameters within the three studied chronological datasets. The work is also motivated generally by an assertion made in a recent study by Lokan and Mendes [14] that "*the use of a window represents more closely what occurs in practice*."

This study retains the definition of BMW as well as the *Bellwether method* utilized in the original study [11] (see Sections 3 and 5, respectively, for details). This approach was taken in order to ensure that the experimental design of this study was as similar as possible to that used in the original study. It should be noted that the current study benchmarks the existing *Bellwether method* (i.e., *SSPM* as used in the original study) with two additional *Bellwether methods* (i.e., *SysSam* and *RandSam* elaborated in Section 5). This study also introduces a guideline for sampling *Bellwethers* for predictive modeling (see Section 8).

Note that the researchers conducting this current study also conducted the original study; thus, this is not an independent replication [14], whereby different researchers undertake and report the respective studies.

The following are the changes made and differences noted between the current and the original study [11]:

1. Even though both studies considered chronological datasets, the current study makes use of an additional source, the Desharnais dataset [27], [28], retrieved from the PROMISE repository [21], whilst the original study [11] made use of the ISBSG[1] and Kitchenham [20] datasets.
2. The original study considered cross-company projects from different organizations, whilst the current study considers projects from a single organization (with the exception of the Desharnais dataset). Projects in the Desharnais dataset are from ten organizations [27] and projects in the Kitchenham dataset are from a single organization [20] made available on the PROMISE repository [21]. Thus, with respect to data homogeneity, we would expect projects from the Kitchenham dataset to be homogeneous as compared to projects from the Desharnais and ISBSG datasets.
3. Projects within the Desharnais dataset have a different age span as compared to projects within the ISBSG and Kitchenham datasets.
4. With respect to size, the ISBSG and Kitchenham datasets have population sizes of 4106 and 145 projects, respectively. The Desharnais dataset comprises of 81 projects, a number that is inconsistent with the evidence found in the original study that *Bellwethers* exist in a population size of at least 100 projects. This in itself provides additional motivation for the current study, being an opportunity to investigate whether such exemplary projects (*Bellwether*) exist in a population comprising fewer projects.
5. The original study made use of four weighting functions (Triangular, Epanechnikov, Gaussian, and Rectangular) as has been used in previous studies [12], [15], [19]. The current study makes use of two additional weighting functions, namely Biweight and Triweight [29], [30] to investigate their potential effect on prediction accuracy of the moving window.
6. The original study considered the use of *log transformation* and *z-score normalization* for scaling the data prior to predictive modeling. In this study, we consider a more robust transformation approach as recommended by Kitchenham *et al*. [31] and Wilcox and Keselman [32], namely *trimming* in addition to *log transformation*.
7. The original study employed the Glass $\Delta$ effect size [33] for investigating the practical significance of the BMW. In this study, we make use of a robust effect size measure recommended by Kitchenham *et al*. [31], namely Cliff's $\delta$ effect size [34].
8. With regard to statistical significance, the original study employed the Kruskal–Wallis *H*-test and the robust Welch's *t*-test to evaluate the differences between predictions during modeling. However, Kitchenham *et al*. [31] have shown that for large sample approximation, the Kruskal–Wallis test is affected when variances from the treatment groups are unequal regardless of equal sample sizes. In this study, we employ Brunner's method for multiple comparison and Yuen's test [35] as recommended by Kitchenham *et al*. [31] for pairwise assessment of significant differences between the BMW and the *GP*, irrespective of unequal variances.
9. Three learners were developed in the original study, namely an SEP baseline model (ATLM [24]), a model derived using ordinary least squares regression [26] and another developed using DNN [25]. In this study, however, we employ the DNN (which proved superior in the original study) for our empirical analysis. We argue that if the DNN can be used to learn from these datasets, then other SEP models can do the same, given that the *Bellwether effect* exists in the studied datasets.
10. Three evaluation measures, namely MAE, MBRE and MIBRE were used in the original study for assessing prediction accuracy of the learners. In this study, we employ the MAE only, recommended by Shepperd and MacDonell [33] to be unbiased to underestimation and overestimation. Similarly, Foss *et al*. [36] recommend MAE as an effective, reliable, and unbiased evaluation measure. MAE was used in a recent study by Lokan and Mendes [14] as an effective evaluation measure for assessing prediction model performance.

## 3. PRELIMINARIES

### A. Concept of Bellwether
According to the Oxford dictionary, *Bellwether* refers to a sheep with a bell around the neck that leads a flock.

---
[1] http://www.isbsg.org

The *Bellwether* concept in the domain of software defect prediction is defined by Krishna *et al.* [6] as a simple transfer learning approach consisting mainly of the *Bellwether effect* and the *Bellwether method*. In the domain of SEP, Mensah *et al.* [11] investigated the existence of *Bellwethers* in chronological datasets and defined the notion of a *Bellwether moving window*.

The Bellwether effect states that, given a set of projects from a given repository, there exists a subset of exemplary projects that can form the Bellwether, and such projects considered as the training set yield a relatively high prediction accuracy for the remaining projects. This Bellwether can be considered for predicting the target of a new project.

The Bellwether method uses a step-by-step heuristic approach to search for that particular Bellwether from the given dataset of historical projects and applies it to a new project whose target is to be estimated. Chen *et al.* [8], [9] define this approach as Bellwether analysis.

The Bellwether moving window refers to exemplary and recently completed projects that yield improved prediction accuracy over new project(s) to be estimated [11]. It should be noted that projects forming the Bellwether moving window have their completion dates prior to the starting dates of new projects.

Within computer science the concept of *Bellwether* was first considered by Chen *et al.* [8], [9] in the context of online analytical processing queries. Their intent was to find a small subset of queries that could be used to make accurate predictions of the target of a new query. Krishna *et al.* [6] in the domain of defect prediction implemented the concept to find exemplary projects (referred to as *Bellwether*) from a set of nonchronological projects to successfully make predictions on the remaining projects. Each project was considered as a potential *Bellwether* in each iteration of prediction of the remaining projects until the project data with the best prediction accuracy was obtained. That particular project was considered as the *Bellwether* to make successful prediction of defects in a *new* project (set as a hold-out). Results from these studies [6], [8], [9] show that *Bellwethers* are not rare and can be used for building prediction models for new project cases with relatively higher prediction accuracy than would be achieved otherwise.

This paper investigates the feasibility of using *Bellwether*s in a *moving window* for building effective SEP models. We present a general overview of how the selection of exemplary projects to be used as the BMW for predicting the targets (software effort) of *new* project cases (set as hold-out) in Fig. 1. We first sort a set of *N* projects chronologically and apply a statistical stratification technique to obtain only the recently completed projects. We then check for the existence of *Bellwethers* based on defined *Bellwether methods*, as elaborated in Section 5. We define the moving window selected as BMW with relatively higher prediction accuracy. The potential BMW, $w_i$ is used to predict the effort of projects in each of the remaining windows, $w_j$ in $N \forall i \neq j$. If the prediction accuracy of projects within the remaining windows, $w_j$ is relatively low, then the projects within the potential BMW, $w_i$ are adjusted by reducing or adding more project(s) sequentially until the best relative prediction accuracy is obtained. In each iteration process of updating the potential window, we generate a TPM and verify its stationarity by computing the respective ergodic Markov chain. Further elaboration of the *Bellwether method* in Fig. 1 for selecting the BMW is detailed in Section 5.

**B. Moving Window and Growing Portfolio**

The development of SEP models using recently completed projects with completion dates prior to the start date of a *new* project (regarded as the *moving window*) has been a recent focus of researchers [15], [12], [13], [18]. The performance accuracies of most of these models were found to be superior to those of conventional SEP models. The moving window approach is different from the conventional approach which uses the entire collection of historical projects as the training set for setting up prediction models [37], [24], [38], [5], [39]. *GP* is the term that describes the use of all historical projects (with completion dates prior to the start date of any *new* project) for building a prediction model. Weighting functions have been shown to further enhance prediction accuracy in a recent replication study by Amasaki and Lokan [12]. They applied four functions, namely triangular, epanechnikov, Gaussian, and rectangular (or uniform) as shown in Fig. 2 (noting that application of the rectangular function is equivalent to using unweighted moving window [12]). This current study incorporates the weighted/unweighted functions depicted in Fig. 2 to both the moving window and the *GP*. We also explore the effect of two additional weighting functions, namely the biweight and triweight [29], [30]. Here, projects are weighted in a nondecreasing order following a similar approach to that used by Amasaki and Lokan [12], whereby a more recently completed project has heavier weight. From Fig. 2, $u$ is defined as follows: $u = \frac{x_i}{n}$, $i = 1,2,...,n$. Here, $x_i$ represents project $i$ ranked in a nondecreasing order of completion date in a set of $n$ projects. That is, $u$ decreases as older projects are assigned with lower ranks. The rectangular or uniform weighting function is the simplest of the weighting functions [12]. It takes the value of 1 if the absolute value of $u$ is less than 1. A shortcoming of the rectangular function is that the estimated density function is not smooth and may have some jumps. Alternatively, the Epanechnikov weighting function provides smooth density functions. The rectangular, epanechnikov, triangular, biweight, and triweight functions have finite support, and hence points beyond $u$ units will not add any contribution to the density at such points. On the other hand, the Gaussian function does not support restrictions and hence it takes into consideration all the data values [30]. To the best of our knowledge, this is the first study to consider these two additional weighting functions to investigate the prediction accuracy of *moving windows* and a *GP* in SEP. We evaluate the prediction accuracy performance using the BMW and the *GP* in constructing SEP models. This is done to investigate the feasibility of using a relevant and recently completed subset of chronological projects (*Bellwether moving window*) in building a robust and accurate prediction model.

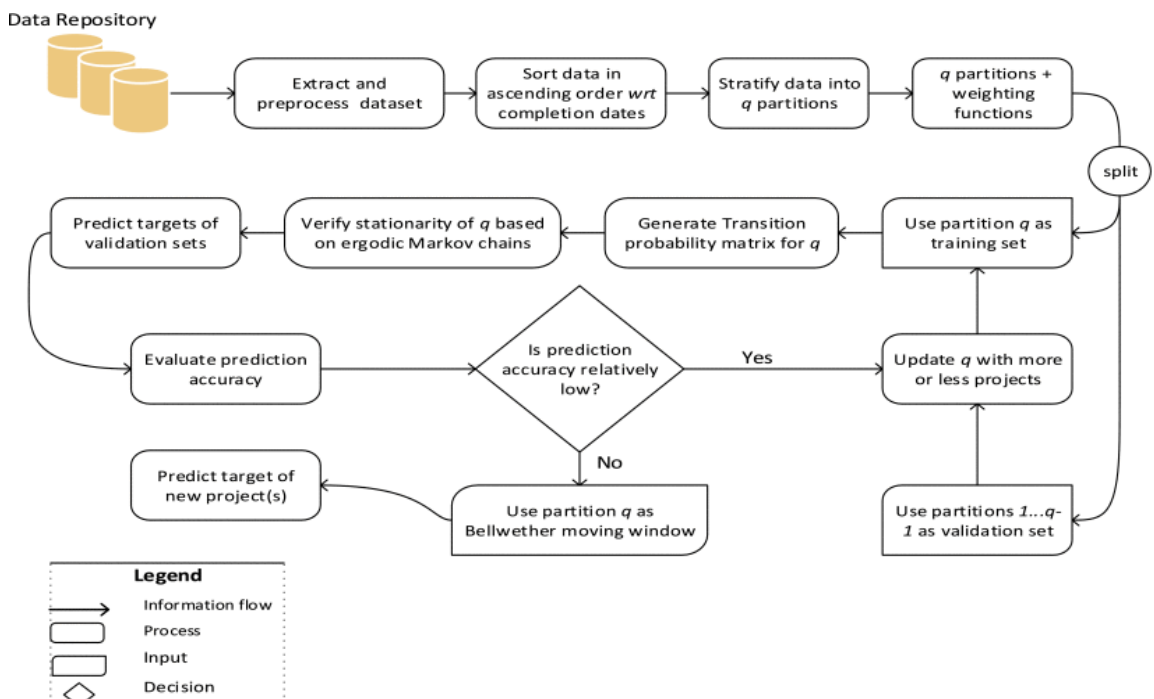

Fig. 1. General overview of the Bellwether method for sampling Bellwether moving windows.

| Weighting function | Formula |
| --- | --- |
| Rectangular (Uniform) | $w(u) = 1$, $|u|<1$ |
| Triangular | $w(u) = 1 - |u|$, $|u|<1$ |
| Epanechnikov | $w(u) = 1 - u^2$, $|u|<1$ |
| Gaussian | $w(u) = exp(-(2.5u)^2/2)$, $|u|<\infty$ |
| Biweight | $w(u) = 15/16(1 - u^2)^2$, $|u| \leq 1$ |
| Triweight | $w(u) = 35/32(1 - u^2)^3$, $|u| \leq 1$ |

Fig. 2. Weighting functions.

**C. Markov Chain Monte Carlo**
MCMC is a probabilistic technique that seeks to solve the problem of sampling by exploring a given sample space through the construction of an EMC whose limiting distribution is the target distribution [10]. MCMC utilizes Markov chains to effectively simulate a random variable $X$ whose future state is independent of past states given the present state. At any time $t$, we can define a Markov process for a present state (in this case the effort of a new project) based on previous history (i.e., historical projects) by finding the transition probabilities from one project state to the previous state with respect to their ages. Thus, in order to make a prediction of a future state (in our case effort) in a Markov process, there is no need to know the entire history of past events (projects) but knowing the current event can be helpful [10]. If we consider a variable $P$ as a TPM of a Markov chain, then each $p_{ij}^n$ element of $P^n$ is the probability that the Markov chain beginning from state $i$ will transition to state $j$ after $n$ steps. The TPM of a given sample with respective project ages can be generated as the probabilities of occurrences from one state to another. MCMC has been applied and proven successful in different software engineering domains such as software testing [40], image processing [41], as well as applied mathematics, statistical and biomedical engineering.

This paper also provides a statistical investigation for the use of the MCMC methodology to obtain a potential subset of projects to be considered as the *Bellwether* (see Section 4). Let **D** denote a set of projects from a given repository (sample space) with $d$ degree of dimensions (features). Then, $X = (x_1,...,x_d) \in D$ can represent a sample whose limiting distribution is known and can form the *Bellwether*. Each of the data points $x_1,...,x_d$ of the sample is allowed to take a discrete or continuous time value which denotes the project ages already known from the given repository (**D**). The Markov chain is then executed a number of times until the chain converges to its limiting distribution to obtain the target sample subset. The selected sample with its respective $t$ ages and constructed limiting distribution forms the *Bellwether* which can be used as the moving window. We define this type of moving window as the BMW.

**D. Central Limit Theorem and Law of Large Numbers**
MCMC relies on the fact that the limiting properties of ergodic Markov chains possess some similarities of independent and identically distributed (*iid*) sequences [10]. Hence, in order to prove the existence of a *Bellwether* sample from a state space, the central limit theorem and the law of large numbers should hold. The central limit theorem states that as the sample size turns to infinity, the sample mean will follow a normal distribution with mean $\mu$ and variance $\sigma^2$ [42]. The law of large numbers states that as the sample size increases, its mean $\bar{x}$ will be approximately the same as the population mean $\mu$ [43]. The larger the sample size ($n > 30$), the smaller the sample variance, $s^2$ and turns to follow a normal distribution [41]. Although Markov chains are not independent sequences, the law of large numbers and central limit theorem are necessary for their independence. Thus, for ergodic Markov chains, successive transitions between visits to the same states are independent [10]. Thus, a moving window sample, $X_i$ can be used as a *Bellwether* with defined window size and window age for the prediction of *new* project cases.

E. Statistical Measures

In order to investigate the statistical significance of *Bellwethers* within the studied datasets, we employ the following robust and nonparametric statistical measures recommended by Kitchenham *et al.* [31].

*1) Brunner's test:* Brunner *et al.* [44] developed a rank version of the analysis of variance (ANOVA) method for testing the statistical significance of treatment effects. This method makes use of midranks for duplicate observations within the same treatment group. Thus, when there are two or more duplicates within a group, the midranks play a key role by allocating the mean of the duplicates to each observation. Brunner's method is relatively reliable and robust to complex factorial designs with unequal variances across treatment groups [31]. Tian and Wilcox [45] found in a recent study that Brunner's ANOVA-type method is more advantageous in terms of Type I error and power as compared to the Agresti–Pendergast method [45]. Brunner's test statistic was recently used in a study by Phannachitta *et al.* [46] for testing the null hypothesis that there is no statistical significance of the difference between predicted effort from two analogy-based estimation variants (i.e., $p > 0.05$). We consider this test statistic because of its ability to perform relatively better when considering small sample sizes for assessing the statistical significance of treatment effects [31], [44], [45]. In this study, we implement Brunner's test statistic to test the null hypothesis that *there are no statistically significant differences across the six weighted/unweighted BMWs and the GP*. Statistical tests are performed at the 5% asymptotic significance level.

*2) Yuen's test:* Given two vectors of data samples with unequal sizes and unequal variances, Yuen's test statistic provides an avenue for robust pairwise testing based on their trimmed means [31], [47]. This robust test statistic was proposed by Yuen [35]. If no trimming is applied to the data, then the test reduces to Welch's *t*-test statistic [48]. Yuen's test uses the trimmed mean as a measure of central location when testing for significant differences between two samples. The test statistic can also be extended to cater for multiple group comparison (i.e., more than two treatment groups). Yuen's test works best when addressing statistically significant differences between groups of different sizes. Hence, we use Yuen's test to determine whether there are statistically significant differences between the *GP* and the BMW. Note that the BMW obtained by the *Bellwether method* [23] is a subset of the *GP* (all historical data considered as training set). Thus, the sample sizes of the two treatment groups are different. Here, we perform this statistical test at the 5% asymptotic significance level.

*3) Cliff's δ Effect Size:* A robust effect size measure as recommended by Kitchenham *et al.* [31], Cliff's delta ($\delta$) is used to find the practical significance of any differences across the BMW and the *GP*. Cliff's $\delta$ effect size is chosen since it yields an effective computation measure irrespective of both the experimental and control groups having different sample sizes. Again, it is not affected by outliers and does not assume that the sampled data follows any particular distribution [49]. The rationale behind our use of Cliff's $\delta$ effect size is that, given two groups of observations that do not necessarily following the same distribution [31], Cliff's $\delta$ effect size is able to determine the amount of overlap that exists between these two groups. This nonparametric effect size measure was also recommended by Arcuri and Briand [50] for empirical software engineering data analysis. Effect size computation is very important since it provides an appropriate measure of the practical usefulness of an experimental effect regardless of the *p-value* showing the statistical significance for a test statistic [31], [50]. Cliff's $\delta$ effect size is defined in (1), whereby each observation $y_i$ from the BMW is compared to each observation $y_j$ in the *GP*. As defined in (1), we count the total number of ($y_i > y_j$) and ($y_i < y_j$), compute the difference and divide by the product of the sample sizes for the prediction observations from the BMW (*m*) and the *GP* (*n*)

$$\delta = \frac{\text{sum}(y_i > y_j) - \text{sum}(y_i < y_j)}{nm}. \quad (1)$$

We interpret the effect size or practical significance based on the magnitude thresholds of Kampenes *et al.* [51] as follows: Negligible effect size ($\delta < 0.112$), small effect size ($0.112 \leq \delta < 0.276$), medium effect size ($0.276 \leq \delta < 0.427$), and large effect size ($\delta \geq 0.427$). We choose an effect size threshold of medium to large since results are misleading when the effect size is relatively small or negligible [4].

*4) Kernel Density Plot:* One of the most effective means of visualizing the distribution of data is the use of the Kernel density plot [31]. Even though histograms can play an important role in visualizing data distributions, variations can occur in the graph depending on the number of bins used [52]. According to Tukey [52], the aforementioned challenge can be addressed using kernel density estimation. This form of estimation addresses the probability density function of a feature or variable by smoothing the histogram. Kernel density estimation employs the kernel weighting function to ensure that the total area within the bins or total area under the curve equals to one. Thus, the estimator creates a *bump* on each data point and computes the total *bumps* using the following equation:

$$p(x) = \frac{1}{nh} \sum K\left(\frac{x - X_i}{h}\right) \quad (2)$$

where *x* is the data point for the density to be estimated; $X_i$ is the midpoint of the interval; *K* is the *bump* or kernel density function; *n* is the sample size; and *h* is the bandwidth [52]. It should be noted that we examine the effect of data transformation to further investigate if it improves the distribution of the data [31].

## 4. POSTULATIONS

In this study, we prove the following four postulations based on theoretical and empirical analysis.

1) *Postulation 1*: If $\Omega$ is a sample space of a set of chronological projects, then the prediction probabilities of its partition sample sets $\{X_1, \ldots, X_q\}$ are independent.

*Proof:* Suppose the sample space, $X$ is partitioned into $q$ partition sample sets. where each $X_i \in \Omega$ is drawn independently, then the prediction probabilities[2] of two events, $X_i$ and $X_j$ can be defined as follows.

Assume that the prediction probability of event Xi occurs after event $X_j$ has occurred (3) and vice versa (4), then according to the conditional probability rule

$$P(X_i|X_j) = \frac{P(X_i \cap X_j)}{P(X_j)}$$
$$P(X_i|X_j) P(X_j) = P(X_i \cap X_j) \quad (3)$$
$$P(X_j|X_i) = \frac{P(X_i \cap X_j)}{P(X_i)}$$
$$P(X_j|X_i) P(X_i) = P(X_i \cap X_j) \quad (4)$$

where $P(X_i) > 0$ and $P(X_j) > 0$.

To show that the events $X_i$ and $X_j$ are independent, then the prediction probability of Xi does not affect the prediction probability of $X_j$ and vice versa. Thus, $P(X_i|X_j) = P(X_i)$ and $P(X_j|X_i) = P(X_j)$.

From the multiplication rule which states that the probability that two events $X_i$ and $X_j$ occur is the product of the probability of $X_i$ and the probability of $X_j$. That is

$$P(X_i \cap X_j) = P(X_i) P(X_j) \quad (5)$$

Substituting (5) into (3) gives

$$P(X_i|X_j) = \frac{P(X_i) P(X_j)}{P(X_j)} = P(X_i). \quad (6)$$

Similarly, substituting (5) into (4) gives

$$P(X_j|X_i) = \frac{P(X_i) P(X_j)}{P(X_i)} = P(X_j). \quad (7)$$

From (6) and (7) it is confirmed that events $X_i$ and $X_j$ from a given sample space, $\Omega$ are independent. For example, $P(X_i|X_j) = P(X_i)$ means that the prediction probability of the occurrence of event $X_i$ given that event $X_j$ has already occurred is the same as the prediction probability of event $X_i$. This implies that $P(X_i \cap X_j) = P(X_i)P(X_j)$ since $P(X_i|X_j) = P(X_i)$ as shown in (6).

Generally, $P(X_i \cap ... \cap X_q) = P(X_i) ... P(X_q)$ which implies that the prediction probabilities of the partition samples, $X_i \in \Omega$ are independent of each other.

We validate this postulation based on empirical analysis, whereby each partition from the sample space resulted in different prediction probabilities of the target variable (software effort). For example, in order to predict the software effort $(Y)$ of pi projects using three partition samples (training sets), $X_1$, $X_2$, and $X_3$ from the ISBSG dataset, we found that their respective prediction probabilities are as follows: $P(X_1) = 0.58$, $P(X_2) = 0.43$, and $P(X_3) = 0.69$. Thus, the training set, X1 can predict about 60% of the actual effort of the $p_i$ projects, $X_2$ can predict about 40%, whilst $X_3$ can predict about 70% when such samples are used for building prediction models.

2) *Postulation 2*: Given a set of $N$ chronological projects with a finite mean $(\mu)$ and variance $(\sigma^2)$, there exists a potential Bellwether that can be used as a moving window.

*Proof:* Let $\{X^{(1)}, ..., X^{(t)}\}$ be iid random samples chronologically drawn from a given population, $N$. If a random sample $X^{(i)} > 30$ is drawn from $N$, then 1) according to the central limit theorem [42] and the law of large numbers [43], $X^{(i)}$ will follow the normal distribution with mean $\mu$ and variance, $\sigma^2$, that is, $X^{(i)} \sim N(\mu, \sigma^2)$ and 2) the $X^{(i)}$ with the minimum accuracy measure(s) to be considered as the Bellwether can then be used to successively predict the remaining samples, $X^{(j)} \forall j \neq i$. That is, the $i$th Bellwether sample predicts the remaining $j$th samples.

In order to confirm the existence of a potential *Bellwether* from a given population set, $N$ empirical analysis was performed based on statistical stratification of $N$ using the *X-means* clustering algorithm [53], and samples were drawn from each stratum. We realized that the best population stratum from which potential *Bellwether* samples can be drawn was of size not less than 100 projects. This was empirically validated by applying the *Bellwether methods* as recorded in Section 5.

3) *Postulation 3*: If "*P*" denotes a regular TPM of a Markov chain, then there exists an EMC, whose respective window can be used as the Bellwether.

*Proof:* We define the Markov chain as a set of random variables $\{X^{(1)}, ..., X^{(t)}\}$ or a collection of stochastic events $\{X(t)|t \geq 0\}$, whereby given the present event of a state at time $t$, the prediction of the future event at $t + 1$ is independent of the past events. The iid sample space $\{X^{(1)}, ..., X^{(t)}\}$ for all states at their respective times can be described as a Markov chain if

$$P(X_{t+1} = i_{t+1}|X_t = i_t, X_{t-1} = i_{t-1} ... X_1 = i_1, X_0 = i_0)$$
$$= P(X_{t+1} = i_{t+1}|X_t = i_t) = p_{ij}^{(k)}. \quad (8)$$

Assume there exists a variable "*P*" that can be formulated as a matrix of transition probabilities of a Markov chain [10] from the sample space $\{X^{(1)}, ..., X^{(t)}\}$, where $t \in T$ denotes the project ages, then the $ijth$ element, $p_{ij}^{(k)} \in P$ is the probability that the Markov chain starting from a particular state, $t_i$ will transition to $t_j$ after $k$ steps. If $p_{ij}^{(k)}$ is homogeneous (or regular), then (9) holds and there exists a unique probability matrix, $\theta_u$ that can form the EMC such that for any $\theta_o$ and for large values of $u$, (10) can be defined

$$P(X^{(k)} = j|X^{(k-1)} = i) = p_{ij}^{(k)} \quad (9)$$
$$\lim_{n \to \infty} \theta_{u+1} = P^u \theta_1 \quad (10)$$

---
[2] Prediction probability is the ratio of the prediction outcome to the actual outcome.

Thus, the EMC can be obtained given a nonnegative power of "$P$" by making all entries of the probability matrix nonzero and irreducible (i.e., systematically squaring $P$). If the limiting state probability matrix (EMC) exists from the TPM constructed with the moving window sample, then that particular sample can be used as the *Bellwether*. We consider such a moving window sample as *Bellwether* since its limiting or stationary distribution has been reached and hence it can form a potential training set for modeling given it has the best prediction accuracy (see details in Section 7).

4) *Postulation 4:* Given a Bellwether whose ergodic Markov chain is known, then its size and age can be defined.

*Proof:* Let a sample $X$ with projects $\{p_1, \ldots, p_n\}$ be classified into $t_i$ states based on their respective project ages. Assume the projects $\{p_1, \ldots, p_n\}$ from a particular sample space are sorted in nondecreasing order based on the completion dates of projects and the EMC of the sample is known. Then, the age of the *Bellwether* can be found as the difference between the maximum and minimum project ages of the sample whose EMC is known. Similarly, the size of the *Bellwether* can be found as the total number of projects within that particular sample (see details in Section 7).

## 5. BELLWETHER SAMPLING METHODS

In order to support the possible replication of this study, we now describe the *Bellwether methods* used to obtain the *Bellwethers* from chronological datasets in this section. That is, we provide detailed step-by-step procedures to find the existence of *Bellwethers* and how they can be used in training a prediction model for *new* projects. Three *Bellwether methods,* namely *SSPM, SysSam,* and *RandSam* were used to sample the *Bellwether* projects from the three studied datasets.

### A. Stratification, Sampling, and Prediction Method (SSPM)

*SSPM* operates based on a probabilistic approach, namely MCMC [10] to sample exemplary projects from a given dataset for prediction purposes. Given a *new* project with unknown software effort, select exemplary projects to be considered as the BMW from a historical dataset ($D$) using the following three main operators (*SORT+CLUSTER, GENERATE TPM,* and *APPLY*):

*SORT + CLUSTER*

*Given a set of N historical and completed projects from D, sort based on the project completion dates and stratify the data into q clusters.*

1. For all $N$ projects from $D$, sort in increasing order using their respective project completion dates.
2. Apply the *X-means*[3] clustering algorithm [53] to the sorted data to obtain $q$ clusters. Perform data stratification based on the $q$ clusters obtained. That is, stratify $N$ into $q$ clusters, whereby each cluster (or window) has a set of chronologically arranged projects.
3. For each cluster or window, compute the weighted moving window by applying the respective weighting functions on all the $q$ windows (see Fig. 2 in Section III). We define the resulting $q$ weighted windows in ascending order (with respect to project age) as $w_1, w_2 \ldots w_q$.
4. Use $w_q$ as a baseline window. It should be noted that, $w_q$ contains recently completed projects in a chronological order and so can form a *baseline weighted moving window*.

*GENERATE TPM*

*Generate the TPM for the resulting weighted moving window and find the respective ergodic Markov chain.*

1. For the resulting baseline window ($w_q$), generate a TPM of the Markov chain following an approach similar to that reported by Dobrow [10]. Here, we use the project ages in $w_q$ (as transition states) with their respective targets (effort) to generate the TPM. The sum of the $p_{ij}$ in each row of the TPM is approximately equal to 1.
2. Compute the EMC for the generated TPM in order to validate the limiting (or stationary) distribution of $w_q$. Thus, using the TPM with $t_i$ states, successfully perform the squaring of TPM until TPM approaches the limiting or stationary distribution. That is, when individual probability elements, $p_{ij}$ of TPM are positive and cannot be reduced further, we say that TPM is ergodic and regular (stationary) [10].
3. Report $w_q^*$ as a BMW if the weighted moving window, $w_q$ is stationary and provides the best prediction accuracy for the majority of the remaining windows, $w_1, w_2 \ldots w_{q-1}$. Else go to step 4 to update $w_q$ with additional project(s) from $w_{q-1}$ and repeat steps 5 and 6. Conversely, update $w_q$ by sequentially removing project(s) from $w_q$ and adding to $w_{q-1}$ and repeating steps 5 and 6.

*APPLY*

*Apply the BMW, $w_q^*$ to the new project data whose software effort is to be predicted.*

1. Once $w_q^*$ is obtained as a BMW in step 7 prior to its application to the *new* pi project data (set as a hold-out), its size and age can be defined from its dimensions. The size of the moving window is defined as the total number of projects within the BMW whilst the age of the moving window is defined as how old the projects within the BMW has been relative to the pi project(s).
2. Predict the software effort of the pi project(s) using the resulting BMW.

### B. Benchmark Sampling Methods
Aside from using the *SSPM* to sample *Bellwether* projects, we further explored two additional benchmark sampling methods. These are the *systematic* and *random sampling* methods. Note that the sampled *Bellwethers* from the three *Bellwether* methods ( *SSPM, systematic,* and *random*) were all compared against the *GP* for setting up the prediction model.

---
[3] *X-means* automatically estimate the number of optimal clusters with respect to the Bayesian Information Criterion [53].

*1) Systematic Sampling (SysSam)*
This type of sampling method refers to the sequential selection of projects from the sorted *GP* given that the selected projects have the most recent ages. Thus, the sampled projects forming the *moving window* are selected from recently completed projects to the oldest completed projects. For example, given that *N* projects are sorted chronologically (i.e., arranged in ascending order of project completion dates), then systematic sampling (*SysSam*) deals with the sequential selection of recent project cases from bottom to top of *N*. A simple *for loop* is implemented whereby each sequential $m_i$ subset is considered as the training set (potential *Bellwether*) and tested with the remaining $N - m_i$ projects (validation set). In each iteration phase, the MAE is used to assess the prediction performance of each sampled subset. The subset with the best prediction performance (minimum MAE) is used to predict the $p_i$ projects set as hold out. Here, we consider the subset with the best prediction performance as the *Bellwether*.

*2) Random Sampling (RandSam)*
Random sampling (*RandSam*) refers to the stochastic selection of projects without necessarily following a definite order. Here, a random sampling algorithm is implemented to select project subsets from the *GP*. The project subsets selected at each iteration did not necessarily contain recently completed projects or have the same sample sizes. Here, each $m_i$ project subset is considered as a training set and is used to predict the targets of the remaining $N - m_i$ projects, considered as the validation set. Following a similar procedure as used in the *SysSam,* we assess the performance of the random subset selection with MAE and consider the subset with the best prediction accuracy as the *Bellwether*. The *Bellwether* is used to make predictions of the pi projects. The *Bellwether* sampled by *RandSam* is benchmarked against the *Bellwethers* sampled by *SSPM* and *SysSam* as well as the *Gp*.

C. Assumptions
The following assumptions are made regarding the effective functioning of the *Bellwether methods*:

1. Each project used in modeling has a start and completion date. Thus, we pruned off nondated projects prior to modeling.
2. All sampled projects from *N* have the same set of features.
3. Each $p_{ij}$ element in the TPM lies within 0 and 1. That is, $0 \leq p_{ij} \leq 1$.
4. The sum of each ith row of the TPM should be approximately 1. That is, $\sum p_{ij} = 1$.
5. All entries of the ergodic Markov chain are nonzero.

Table I. DESCRIPTIVE STATISTICS OF SAMPLED FEATURES IN THE DESHARNAIS DATASET

| Feature | N | Min | Max | Mean | Std.Dev | Skew | Kurt |
|---|---|---|---|---|---|---|---|
| TeamExp | 79 | 0 | 4 | 2.27 | 1.34 | -0.042 | -1.26 |
| ManagerExp | 78 | 0 | 7 | 2.67 | 1.52 | 0.20 | 0.07 |
| Transactions | 81 | 9 | 886 | 179.90 | 143.32 | 2.36 | 7.73 |
| Entities | 81 | 7 | 387 | 122.33 | 84.88 | 1.34 | 1.48 |
| PointsAjust | 81 | 73 | 1127 | 302.23 | 179.68 | 1.78 | 4.94 |
| Envergure | 81 | 5 | 52 | 27.63 | 10.59 | -0.11 | -0.28 |
| PointsNonAjust | 81 | 62 | 1116 | 287.05 | 185.11 | 1.67 | 4.16 |
| Effort | 81 | 546 | 23940 | 5046.31 | 4418.77 | 2.01 | 4.72 |

Min and Max denote Minimum and Maximum, respectively.
Std.Dev denotes standard deviation.
Skew and Kurt denote skewness and kurtosis, respectively.

## 6. EMPIRICAL FRAMEWORK AND METHODOLOGY

A. Dataset Descriptions
*1) Desharnais Dataset:* This dataset was collected by Desharnais [27] from ten Canadian software organizations and it has been made publicly available at the PROMISE repository [21] . It is made up of a total of 81 management information system development projects with ten features. Projects in the dataset were completed between 1983 and 1988 over an approximate span of 6 years. Following a similar procedure by Li *et al.* [54], [55] and Shepperd and Schofield [28], eight features were selected for modeling. The independent features are *team experience (teamExp), manager's experience (managerExp), programming language, number of entities, number of unadjusted function points (UFP, pointsNonAdjust), number of transactions,* and *the FPA technical complexity scale factor (envergure).* The dependent feature is the development *effort* variable measured in person-hours. The remaining two features within the Desharnais dataset are the *length* (or *duration*) and the *number of adjusted function points (AFP, pointsAdjust).* With the exception of the *language, teamExp,* and *managerExp* features which are categorical variables, the remaining features are numerical variables. Note that this study considered *prior features* (i.e., independent variables known prior to project development) for setting up the SEP models. The *length* or project duration variable was therefore not used as an independent variable in the prediction models because it can only be known at the end of project development. In fact, the *length* variable could be considered as a dependent variable in some cases; hence, it was not used as an independent variable in this study. Note that the *AFP* variable is a *prior feature* that can equally be considered as an independent variable for setting up an SEP model, similar to the case of the *UFP* variable used in our study. The difference between the two variables is that the *AFP* value is a multiplication of the *UFP* by a scale factor, as described by Desharnais [27]. The two variables are strongly correlated (with a correlation coefficient value, $r = 0.986$) and so their inclusion would negatively affect model stability and interpretability if used in the same SEP model [3]. We provide summary statistics of the project features in Table I.

Since the Desharnais dataset does not have the commencement or start date of project development, we computed it by subtracting the development duration or length variable (*length*) from the end date variable ( *yearEnd*). As a result of the *length* variable being in months and the *yearEnd* variable in years [27], we converted *length* to years for uniform computation of *yearStart*. Thus, start date (*yearStart*) is defined in ( 11) as follows:

$$yearStart = yearEnd - length. \quad (11)$$

*2) Kitchenham Dataset:* Kitchenham *et al.* [20] were the first to use a dataset for moving window modeling in SEP. The Kitchenham dataset, which is publicly available at the PROMISE repository [21], includes projects undertaken between 1994 and 1999, with an approximate span of 6 years. It is a chronological dataset comprising 145 maintenance and development projects with 9 features from a single organization, namely the Computer Science Corporation (CSC). Further details about this dataset are provided in previous studies by Kitchenham *et al.* [20] and Tsunoda *et al.* [56]. The features or variables within the dataset include *project type, actual duration, actual start date, estimated completion date, AFPs, client code[4], first estimate, first estimate method,* and *actual effort*. Here, we selected two *prior* features [25], namely *AFPs* and *project type* as the independent variables and *actual effort* as the dependent variable, following a similar procedure to that used in our previous study [11]. We did not use the *posterior* feature (i.e., *actual duration*) as an independent variable since it is impossible to determine the actual duration of a project prior to its development. Note that the *actual duration* variable is measured in days [20] and therefore for consistency, we converted it to years prior to modeling.

The 145 projects within the Kitchenham dataset are distributed across the six clients of the CSC organization as follows: Client 1: 16 projects, Client 2: 116 projects, Client 3: 4 projects, Client 4: 4 projects, Client 5: 1 projects, and Client 6: 4 project. Out of the 145 projects, 3 projects were detected as having missing entries with respect to the *estimated completion date* variable. As a result of that, we computed the completion dates for those three project cases by adding the *actual duration* (in years) to the respective *actual start date*. Table II provides a descriptive summary of sampled features within the dataset.

*3) ISBSG Dataset:* The ISBSG dataset release 10 sourced from the ISBSG repository was used for this study. It is a chronological dataset with a total of 4106 projects and 105 features from over 25 countries. It contains variants of projects from different organizations, such as communications, insurance, banking, manufacturing, government, and business services. All projects are within the range of May 1988 to November 2007 with an approximate span of 20 years. To facilitate an effective replication of existing moving window studies, we used the same ISBSG dataset (release 10) that has previously been considered for prediction modeling [11], [13], [15]. There is no unique pattern with respect to specific programming languages used for development of the projects. Further details of the ISBSG dataset are provided in studies by Li *et al.* [54] and Lokan *et al.* [57].

Table II. DESCRIPTIVE STATISTICS OF SAMPLED FEATURES IN THE KITCHENHAM DATASET

| Feature | Min | Max | Mean | Std.Dev | Skew | Kurt |
|---|---|---|---|---|---|---|
| Actual_duration | 0.1 | 2.6 | 0.57 | 0.37 | 1.93 | 6.12 |
| AFP | 15.36 | 18137.48 | 527.67 | 1521.99 | 10.92 | 126.70 |
| First_estimate | 121 | 79870 | 2855.97 | 6789.29 | 10.34 | 116.91 |
| Actual_effort | 219 | 113930 | 3113.12 | 9598.01 | 10.87 | 125.64 |

*AFP denotes adjusted function points.*

Table III. DESCRIPTIVE STATISTICS OF SAMPLED FEATURES IN THE ISBSG DATASET

| Feature | N | Min | Max | Mean | Std.Dev | Skew | Kurt |
|---|---|---|---|---|---|---|---|
| AFP | 614 | 3 | 4911 | 290.06 | 466.49 | 5.55 | 43.72 |
| Norm_PDR | 606 | 0.3 | 315.6 | 16.49 | 19.97 | 7.48 | 94.39 |
| Team_size | 260 | 1 | 50 | 9.12 | 8.79 | 2.16 | 5.48 |
| Input_count | 269 | 0 | 2075 | 97.15 | 182.09 | 7.05 | 65.35 |
| Output_count | 268 | 0 | 667 | 67.60 | 109.58 | 3.06 | 10.84 |
| Enquiry_count | 269 | 0 | 639 | 43.50 | 72.95 | 4.21 | 25.16 |
| File_count | 268 | 0 | 1208 | 64.71 | 124.75 | 5.05 | 33.57 |
| Interface_count | 265 | 0 | 747 | 20.09 | 75.43 | 6.91 | 53.97 |
| Added_count | 277 | 0 | 3738 | 182.73 | 375.78 | 5.10 | 35.62 |
| Changed_count | 276 | 0 | 1369 | 89.30 | 153.95 | 4.28 | 26.07 |
| Deleted_count | 273 | 0 | 2657 | 15.77 | 162.72 | 15.87 | 257.9 |
| Norm_Effort | 640 | 4 | 88555 | 3508.9 | 7020.4 | 7.02 | 63.20 |

*Norm_PDR denotes normalised project delivery rate*

The projects in the ISBSG dataset are submitted by different organizations with different policies and development methodologies. Hence, such cross-organization projects are heterogeneous in nature [54] and do not share common characteristics such as common development team, programming skills, and other development policies. Different from our original study [11] that used cross-organization projects, this study sampled projects from a single organization type for our empirical investigation. Hence, a total of 699 projects (22.1%) from Communication sector organizations have been considered in this study. The selected features from the ISBSG dataset are the *size* of the projects measured in UFP, *language type* (3GL and 4GL), *development type* (new development, redevelopment, and enhancement), *development platform* (PC, mainframe, midrange, and multiplatform), and the normalized *development effort* measured in person-hours [13], [15], [56]. The normalized *effort* is the dependent variable [54], while the remaining *prior* features are the independent variables. We provide the summary statistics of the sampled features in Table III. Project delivery rate (PDR) is a feature in the ISBSG dataset that was examined in a previous study by Amasaki and Lokan [12]. They found that PDR changes with time and low PDR suggests high productivity. This finding supports the use of moving windows in prediction modeling [15].

Although the aforementioned three datasets seem old, the focus is not on how old the datasets are but to investigate the application of the *Bellwether effect* and Markov chain Monte Carlo analysis in SEP modeling.

**B. Data Preprocessing**
*1) Elimination of Irrelevant Instances and Missing Data Points:* A systematic analysis [5] was performed on all data instances in each dataset to address missing data values and project cases that were not relevant to the analysis. A handful of four missing entries were found in the Desharnais dataset and eliminated prior to setting up the prediction model, as has been done in previous studies [3], [28], [54], [55]. Note that the four missing data entries were observed across the selected independent variables. We found three missing data points in the Kitchenham dataset with regard to the *estimated completion date* feature. Since the dataset extracted already has the *actual start date* and the *actual duration* for each project, we estimated the three missing values by adding the *actual duration* to the *actual start date*. Thus, no data point was eliminated from the Kitchenham dataset prior to prediction modeling.

---
[4] The client code denotes the various categories of clients {1, 2, 3, 4, 5, 6} in which the CSC organisation develops or maintains software products on their behalf.

Similar to the original study [11], we eliminated irrelevant project cases in the ISBSG dataset as follows:

1. Eliminate projects if their respective project elapsed times or implementation dates are unknown [13], [15].
2. Eliminate projects with low data quality rating (that is, projects not rated as A or B by ISBSG are excluded).
3. Eliminate projects whose UFP rating is not A as suggested by ISBSG [58].
4. Similar to previous studies [59], [60], [61], select projects based on functional sizing method with IFPUG version 4.0 and above.
5. Eliminate projects with normalized ratios more than 1.2 as recommended by ISBSG [58]. The normalized ratio is defined as the ratio of the normalized effort to the summary effort, suggested by ISBSG for project subset selection and refinement [58], [59].
6. Eliminate projects with unknown development team effort as reported in the ISBSG dataset. That is, projects with resource level "1" denoting development team effort are selected, following similar approach in previous studies [59], [62], [63].
7. Eliminate projects with unknown organization types [60].
8. Eliminate projects that are considered as web projects [13], [15]. This is due to the established fact that web projects have different properties from the other projects in this dataset [13]. Also function point analysis does not reflect all the necessary features that have an influence in the determination of effort of web projects.
9. Eliminate projects with missing data entries based on the selected features.

*2) Outlier and Influential Data Point Detection:* Outliers and influential data points are known to affect the performance of prediction models when left unresolved. Following the stability analysis approach of Mendes and Kitchenham [64], we first fit prediction models without considering data transformations and we investigate the effect of outliers and influential data points. Second, we transform the studied datasets and again investigate the effect of outliers and influential data points, in each case evaluating the performance of the prediction model. Outliers and influential data points are identified and removed provided they negatively affect the performance of the predictive modeling approach [3], [5], [33], [61], [62]. Outliers are identified based on the Cook's distance measure [65] and kernel density plots (recommended by Kitchenham *et al.* [31]). Cook's distance is used following a similar approach by Seo and Bae [63] in the identification and treatment of outliers during model construction. Thus, we considered an observation to be an outlier if its Cook's distance value is at least three times more than the mean Cook's distance.

Table IV. STATISTICS OF OUTLIERS AND INFLUENTIAL DATA POINTS IDENTIFIED AND TRIMMED OFF

| Dataset | Cook's Distance | | DFFITS | | Trimmed off |
|---|---|---|---|---|---|
| | unlog | log | unlog | log | |
| Desharnais | 4 | 3 | 6 | 3 | 1 |
| Kitchenham | 1 | 1 | 3 | 1 | 1 |
| ISBSG | 9 | 2 | 9 | 5 | 3 |

With regard to influential data points, we used the DFFITS diagnostic [66], [67] to determine such observations. We considered an observation to be influential when its exclusion results in major changes in the prediction model. Thus, an observation is considered influential if its DFFITS exceeds 1 for small to medium datasets and $2\sqrt{2/n}$ for large datasets.

*3) Data Trimming + Transformation:* The use of heavy-tailed distributed data for predictive modeling does not provide reliable prediction estimates. It is therefore important to consider necessary data transformation measures to investigate their effect on prediction accuracy. A recent study by Kitchenham *et al.* [31] recommended the use of *trimming* instead of simple transformation techniques (e.g., *log transform*, square root, box cox) as a more robust approach. Simple transformations are known to transform skewed data to (more) symmetric distributions but do not necessarily deal with outliers [33], [34], [64]. Trimming of data was proposed by Wilcox and Keselman [32] to address heavy-tailed distributed data. Trimming is the process of eliminating the tails of a distribution in a population or sample under investigation prior to setting up the prediction model. This results in an average of a trimmed subset (*trimmed mean*) used as a measure of central location. One *trimming* method suggested in previous studies [34], [64] is to remove a fixed proportion of the smallest and largest values given that the data is sorted in order of ascendancy [32]. A 20% threshold was suggested as a plausible proportion to be trimmed. For example, given that a dataset consists of 20 data points, 20% trimming refers to eliminating the 2 smallest and 2 largest data points, assuming the data is sorted. Another *trimming* method suggested by Wilcox and Keselman [32] is to detect and remove outliers from the dataset and compute the average of the remaining data subset.

Both methods alter the dataset prior to setting up a prediction model. In this study, we adopt the second method and eliminate outliers provided they do not play a significant role in the predictive modeling. Afterwards, we conduct empirical analyses with and without a simple transformation technique to further investigate its significant effect. We make use of the *log transformation* [69], which was shown in our original study [11] to improve prediction accuracy. Previous moving window studies [12]–[15] considered the use of *log transformation* to transform features (e.g., size, duration, effort, and so on) to follow the normal distribution. We seek to confirm in this study whether the use of *log transformation* on a subset of *trimmed* data plays an important role in prediction analysis.

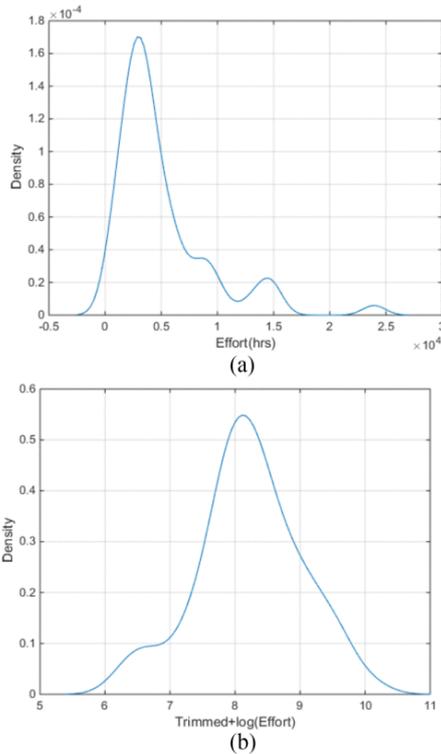

Fig. 3. (a) Kernel density plot for the *untransformed* Desharnais dataset. (b) Kernel density plot for the *trimming + log transformed* Desharnais dataset.

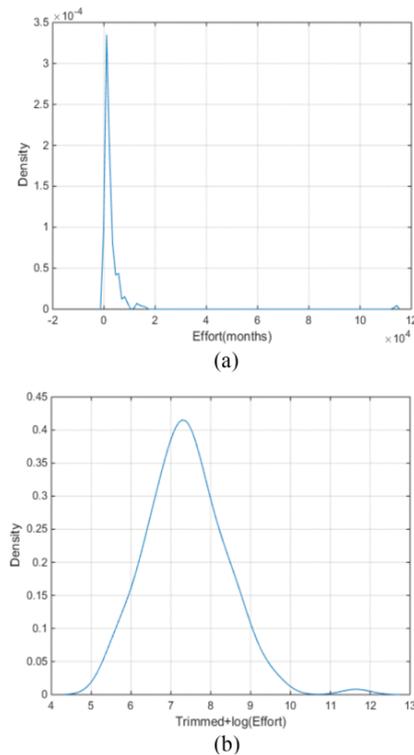

Fig. 4. (a) Kernel density plot for the *untransformed* Kitchenham dataset. (b)

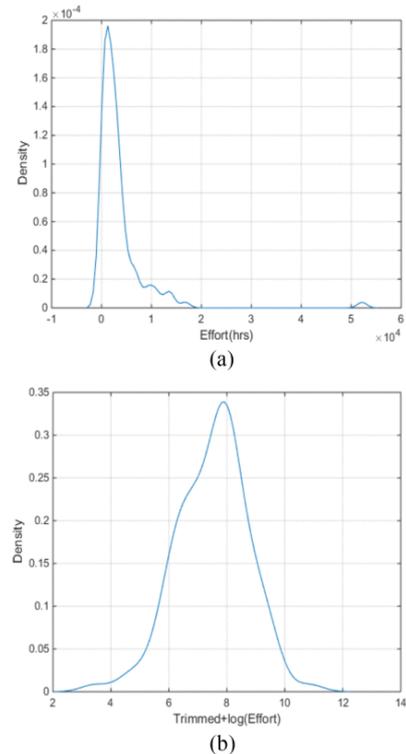

Fig. 5. (a) Kernel density plot for the *untransformed* ISBSG dataset. (b) Kernel density plot for the *trimming + log transformed* ISBSG dataset

For example, in Table 4 three outliers and three influential data points (i.e., the 10th, 19th, and 64th project cases) were identified by the Cook's distance measure as well as the DFFITS in the case of the *log transformed* Desharnais dataset. We found that only the 64th project case was very influential or significantly affected the predictive modeling. Hence, we *trimmed off* that project case prior to investigating the *Bellwether* projects. In the Kitchenham dataset, an outlier and influential data point (i.e., the 62nd project case) was detected by the Cook's distance and DFFITS in the case of the *log transformed* data (see Table 4). This influential data point was removed since it significantly affected the predictive modeling. It should be noted that in the case of the untransformed data, the same 62nd influential data point affected the prediction model. However, in the ISBSG dataset, three influential data points were *trimmed off* (i.e., the 23rd, 26th, and 74th project cases) when the *log transformation* was applied.

The Kernel density plots in Figs. 3– 5(a) and Figs. 3– 5(b) show the distribution of the target variable (*effort*) before and after *trimming + log transformation,* respectively. The distribution of the data after *trimming + log transformation* was observed to follow the normal distribution (b) as compared to before data transformation (a).

In order to make an inference that the existence of *Bellwether* should follow a normal distribution (Postulation 2), we further performed a statistical test of normality using the Shapiro–Wilk test [70]. We found that the selected features followed the normal distribution with $p$-values (>0.05) at 5% asymptotic significance level. Result from the Shapiro–Wilk test of normality confirms that, for a *Bellwether* to yield a relatively superior prediction accuracy, the project data must follow the normal distribution. Note that for normality, the *p-value* should be more than the asymptotic statistical significance level ($\alpha = 0.05$). Thus, the error evaluation measures (MAE) resulted in *p-values* of more than 0.05 for the Shapiro–Wilk test, respectively.

The aforementioned preprocessing procedure resulted in a subset of 117 projects (16.7%) out of the sample space of 699 projects from the Communication organization type in the ISBSG dataset. Out of the preprocessed subset selection, 107 projects (91.5%) were considered as the *GP* and 10 projects (8.5%) with recent completion dates were set aside as hold-out, for testing the prediction accuracy of the deep learning model (DNN). Note that the *Bellwether* was sampled from the *GP* in the empirical analysis conducted in this study. The three *Bellwether methods* (i.e., *SSPM*, *SysSam* and *RandSam*) were used to select the *Bellwether* to be used as the training set for the predictive modeling. Similarly, with regard to the Desharnais dataset, 76 projects (93.8%) resulted from the preprocessing. Out of these 76 preprocessed projects, we used 5 projects (6.6%) with recent completion dates relative to the remaining projects as the hold-out. Thus, we selected the 5 projects that had 1988 as their completion dates. The remaining 71 projects (93.4%) ranging from 1982 to 1987 were considered as the *GP* out of which the *Bellwether* was sampled.

Lastly, with regard to the Kitchenham dataset, 144 projects resulted from preprocessing. Out of this total, 134 projects (93.1%) were considered as the *GP* and 10 projects (6.9%) as the hold-out.

**C. Experimental Setup**

This study uses the *SSPM* approach together with the *SysSam* and *RandSam* methods (see Section 5) to investigate the selection of exemplary projects to be considered as the *Bellwether*. The sampled *Bellwether* (with defined window size and age) from each *Bellwether method* is benchmarked against the *GP* for predicting the targets of $p_i$ hold-out projects. Following the *Bellwether methods* in Section 5, comparison is made across the application of the weighted functions (triangular, epanechnikov, Gaussian, biweight, and triweight) and the unweighted function (rectangular or uniform) on the selection partitions in each dataset. For each sampled partition set, we evaluate its accuracy performance (MAE) by using the sampled partition as the training set and performing successive predictions on the remaining unselected partitions (validation set). Thus, for each iteration, the selected partition set is used as the training set and the unselected partition considered as the validation set until the best partition with the minimum MAE is obtained. The best partition set is considered as the *Bellwether* following the detailed procedure in Section 5. The *Bellwether* benchmarked against the *GP* is used for predicting the targets of the $p_i$ projects. We used leave-one-out cross validation [39], as per our original study [11], to obtain the training and validation sets for the *GP*. Thus, at each run, $N-1$ projects are used for training the prediction model and the remaining project for validation. The best $N-1$ partition set of the *GP* is used to predict the pi projects and benchmarked against the prediction performance of the *Bellwethers*.

*1) Deep Neural Network Model:* The prediction model or learner considered in this study is a DNN, which had resulted in a relatively higher prediction accuracy in our original study as compared to the ATLM [24] and ordinary least squares regression [26]. A deep neural network is a feedforward artificial neural network architecture which has more than one hidden layer with their respective interconnection of neurons. Deep learning models have yielded improved prediction accuracy in a range of previous studies [25], [71] –[76].

We setup a DNN model which makes use of multiple hidden layers and an output layer with their respective neurons to automatically learn from a set of projects and provide the resulting prediction for the target. Thus, given a supervised learning problem with input data ($X_i$) and its respective label ($Y_i$), we setup a DNN model with *h* hidden layers containing a set of neurons and an output layer with a single neuron. The single neuron in the output layer is responsible for *firing* the estimated variable/label (i.e., software effort). Each neuron uses a nonlinear activation function to map the output from one neuron in a given layer to an interconnected neuron in the next layer. We made use of the hyperbolic tangent function [77], [71] as the activation function in each of the neurons for *firing* the respective outputs based on given input values *z*. The hyperbolic tangent function, $tanh(z)$ defined in (12) has a *firing* output range of $(-1,1)$ that prevents the DNN model getting *stuck* during the training process. That is, unlike the sigmoid activation function with an output range of $(0,1)$ that results in an output value close to 0 when the *z* value is negative, $tanh(z)$ is able to output the corresponding negative value for a given *z* due to its wide range of output $(-1,1)$. This makes the use of the $tanh(z)$ an efficient activation function when setting up the DNN:

$$\tanh(z) = \frac{e^z - e^{-z}}{e^z + e^{-z}} = \frac{e^{2z} - 1}{e^{2z} + 1} \quad (12)$$

In order to obtain the best DNN model, we started with a single hidden layer and an output layer each containing a single neuron and iterated through a series of fine-tuning processes to obtain the best model parameters at a minimal loss. At each iteration, the weights $w_{ij}$ connecting the neurons were initialized to 0. The new weights $w_{ij}^{(t+1)}$ were updated after each iteration process using the function defined in (13) whereby the change $\Delta w_{ij}^{(t+1)}$ in weight for each new weight is defined in (14). The parameter *t* denotes the level of iteration, $\alpha$ is the learning rate, $\gamma$ is the momentum, and $L(\cdot)$ is the loss function or the error between the predicted output and the actual label. Note that the Levenberg–Marquardt backpropagation optimization [78], [79] training function was employed to update the weights of the neurons in the hidden and output layers, respectively. This training function, which is a supervised learning algorithm, is relatively efficient [80] when training with small to moderate-sized datasets

$$w_{ij}^{(t+1)} = w_{ij}^{(t)} + \Delta w_{ij}^{(t+1)} \quad (13)$$

$$\Delta w_{ij}^{(t+1)} = \gamma \Delta w_{ij}^{(t)} - \alpha \frac{\partial L(\cdot)}{\partial w_{ij}}. \quad (14)$$

TABLE V. AE EVALUATION OF POTENTIAL BELLWETHERS SAMPLED BY *SSPM, SysSam,* AND *RandSam* AGAINST THE GROWING PORTFOLIO BENCHMARK IN THE DESHARNAIS DATASET

| Window | | Stratification, Sampling and Prediction Method (*SSPM*) | | | | | |
|---|---|---|---|---|---|---|---|
| Age (years) | Size | Rectangular | Triangular | Epanechnikov | Gaussian | Biweight | Triweight |
| 1 | 38 | 1.6926 | 1.2109 | 1.8054 | 0.9257 | 1.0748 | 0.8947 |
| 1.5 | 42 | 0.9914 | 1.5862 | 0.7931 | 0.7035 | 0.8675 | 0.7184 |
| 2 | 46 | 0.8351 | 1.5682 | 0.7841 | 0.9801 | 0.8577 | 0.3170** |
| 2 | 47 | 1.1288 | 1.2563 | 0.8628 | 1.0785 | 0.9437 | 0.7089 |
| 5 | 71(GP) | 2.4018 | 2.9906 | 2.4953 | 1.8691 | 1.6355 | 1.2318 |
| Age (years) | Size | Systematic Sampling Approach (*SysSam*) | | | | | |
| 1.5 | 40 | 1.8126 | 1.5444 | 1.0889 | 0.8056 | 0.9552 | 0.7104 |
| 1.5 | 44 | 1.7410 | 1.4964 | 0.9929 | 1.2055 | 1.4300 | 0.6241** |
| 2 | 45 | 1.9154 | 2.0302 | 1.0604 | 1.2878 | 1.1268 | 0.9658 |
| 2 | 49 | 1.0103 | 1.6904 | 1.3809 | 0.6504 | 0.7552 | 0.8473 |
| 5 | 71(GP) | 2.4018 | 2.9906 | 2.4953 | 1.8691 | 1.6355 | 1.2318 |
| Age (years) | Size | Random Sampling Approach (*RandSam*) | | | | | |
| 1.5 | 36 | 1.9190 | 1.2626 | 1.5251 | 1.3282 | 1.2872 | 1.2462 |
| 2 | 41 | 1.1594 | 1.0678 | 2.1356 | 1.3347 | 1.1680 | 1.0011 |
| 2 | 45 | 1.7626 | 1.2521 | 1.5042 | 0.3151 | 1.2757 | 1.2363 |
| 2 | 48 | 1.1361 | 0.9119 | 1.8239 | 1.1399 | 0.9975 | 0.8549** |
| 5 | 71(GP) | 2.4018 | 2.9906 | 2.4953 | 1.8691 | 1.6355 | 1.2318 |

GP denotes the growing portfolio.
Greyed cells used to indicate best prediction performance across weighted/unweighted windows for each aged/sized sample set. Thus, the minimum MAE signifies the best prediction accuracy.
"**" denotes the best weighted/unweighted window for each sampling approach (SSPM, SysSam and RandSam). That is, the window with the minimum MAE for all greyed cells per each sampling approach.

The choice of the learning rate ($\alpha$) can significantly affect the performance of a DNN during the training process. Thus, there is the need for an optimal selection of $\alpha$ in order for the DNN to converge to relatively better global optima. After commencing the training of the DNN with a relatively large learning rate of $\alpha = 0.1$, we ended up with a relatively low learning rate of $\alpha = 0.01$ which yielded the best performance at 100 training epochs. Note that if the learning rate is too small (e.g., $\alpha = 0.001$), the gradient descent can be slow and training will take a relatively long time to converge. Alternatively, if $\alpha$ is too large (e.g., $\alpha = 0.2$), the gradient descent can wave-off the minimum and training might fail to converge. We recommend training of the DNN to commence with a relatively large learning rate and then gradually decrease it to a relatively lower learning rate to obtain the best prediction accuracy. The momentum ($\gamma$) is used for smoothing out the weight updates in the interconnected neurons. Even though momentum also affects the training process, its impact is less significant relative to the learning rate. The best momentum value used in our DNN was $\gamma = 10^{-6}$ after commencing the training process with a $\gamma = 0.1$ [60].

After iteratively fine tuning the hidden layers with their respective neurons, we found that the best DNN that resulted in improved prediction accuracy $wrt$ to MAE was a two-hidden layered architecture with five and two neurons respectively, in addition to an output layer with a single neuron. Note that the sampled data from each *Bellwether method* was considered as the training set and the remaining unsampled data as the validation set.

### D. Performance Measures
We employ a robust performance measure (i.e., the MAE) which has been proven reliable by Shepperd and MacDonell [33] and Foss *et al.* [36] for assessing SEP models. MAE has been considered in a growing number of previous studies [3], [12], [14], [24], [37], [46] to evaluate the prediction accuracy of SEP models. It is a risk function that measures the average absolute deviation of the estimated effort ($E_E$) values from the actual or true effort ($E_A$) values and it is defined in (15). Minimum values from the performance measures are considered superior in terms of prediction accuracy

$$\text{MAE} = \frac{1}{n}\sum_{i=1}^{n}|E_{Ai} - E_{Ei}|. \quad (15)$$

The robust statistical tests employed in this study for determining the statistical differences among the weighted/unweighted moving windows and *GP* are Yuen's test, Brunner's test, and Cliff's $\delta$ effect size. Statistically significant differences are considered at the $\alpha = 0.05$ asymptotic significance level as well as a confidence interval of 95%. Details of the aforementioned statistical tests are provided in Section 3.

## 7. RESULTS AND DISCUSSION

In this section, we present the empirical results and discussion of the existence of *Bellwethers* against the *GP* benchmark in each of the three studied datasets. Our results are presented based on the three *Bellwether methods* (*SSPM, SysSam,* and *RandSam*) and performance evaluations assessed using MAE, Cliff's $\delta$ effect size, Yuen's test, and Brunner's test at 5% asymptotic significance level.

### A. Bellwether Effect in Chronological Datasets
*1) Desharnais Dataset:* Results obtained after applying the *SSPM* show that, after subjecting the sorted chronological dataset to the *X-means* clustering algorithm, three partitions (with initial approximate sizes of 41, 23, and 8 projects, respectively) were recorded. These initial partitions were used to predict the targets of the $p_i$ projects in order to validate Postulation 1. We found that the prediction probabilities of the three partitions with respect to the $p_i$ projects were different and independent from each other. In order to empirically investigate the existence of a *Bellwether effect* in the Desharnais dataset, we followed the *SSPM* procedure as well as the *SysSam* and the *RandSam* methods as elaborated in Section 5.

TABLE VI. MAE EVALUATION OF POTENTIAL BELLWETHERS SAMPLED BY *SSPM, SysSam,* AND *RandSam* AGAINST THE GROWING PORTFOLIO BENCHMARK IN THE KITCHENHAM DATASET

| Window | | Stratification, Sampling and Prediction Method (*SSPM*) | | | | | |
|---|---|---|---|---|---|---|---|
| Age (years) | Size | Rectangular | Triangular | Epanechnikov | Gaussian | Biweight | Triweight |
| 1.5 | 76 | 1.2534 | 1.3421 | 0.8437 | 0.7666 | 1.0742 | 0.8207 |
| 2 | 83 | 1.0532 | 1.0882 | 0.7637 | 0.9102 | 1.0965 | 0.5267** |
| 2 | 86 | 1.1006 | 1.0430 | 0.8617 | 0.8147 | 0.8521 | 0.7216 |
| 2 | 87 | 0.9507 | 1.2288 | 1.4577 | 0.8604 | 1.2503 | 0.6453 |
| 3 | 134(GP) | 3.2410 | 2.5300 | 2.0601 | 1.6625 | 2.5797 | 2.4969 |
| Age (years) | Size | Systematic Sampling Approach (*SysSam*) | | | | | |
| 2 | 80 | 1.3978 | 1.5370 | 1.0740 | 1.2607 | 1.0874 | 0.9034 |
| 2 | 82 | 1.4591 | 1.6198 | 1.2396 | 1.1748 | 1.1779 | 1.0811 |
| 2 | 84 | 1.6481 | 1.8749 | 1.7497 | 1.0936 | 0.9569 | 0.8202** |
| 2 | 89 | 1.8680 | 1.1718 | 1.3436 | 1.4647 | 1.2817 | 1.0985 |
| 3 | 134(GP) | 3.2410 | 2.5300 | 2.0601 | 1.6625 | 2.5797 | 2.4969 |
| Age (years) | Size | Random Sampling Approach (*RandSam*) | | | | | |
| 2 | 80 | 2.6738 | 1.5264 | 2.0529 | 1.4080 | 1.6696 | 1.3103 |
| 2 | 81 | 1.6198 | 0.7169 | 1.4339 | 0.8962 | 0.7842 | 0.6721** |
| 2 | 82 | 1.8205 | 1.5922 | 2.1845 | 1.9903 | 1.7416 | 1.4927 |
| 2 | 83 | 2.1799 | 2.7044 | 2.4089 | 1.3054 | 1.8643 | 1.5979 |
| 3 | 134(GP) | 3.2410 | 2.5300 | 2.0601 | 1.6625 | 2.5797 | 2.4969 |

The partition containing 41 recently completed projects relative to partitions 2 and 3 was used as the initial baseline window (in the case of *SSPM*). In each iteration, the window size was updated and the respective TPM computed. We assessed the prediction performance using MAE for each updated window in each iteration. Prediction performance (MAE) results of the stationary and potential *Bellwethers* with known ergodic Markov chains are reported in Table 5. Similarly, the best prediction performance results of the sampled *Bellwethers* from the *SysSam* and *RandSam* methods are also noted in Table 5.

Results of the weighted/unweighted windows benchmarked against the GP which yielded the best prediction accuracy for each sampling approach are summarized in Table 5. For ease of understanding of the summarized results, we used shaded or *greyed cells* to denote the best prediction window across the six weighting/unweighting functions. Note that comparisons are made across the window ages and sizes for each sampling approach (*SSPM, SysSam,* and *RandSam*). Lastly, we used "**" to further denote the window which can act as the *Bellwether* in each of the sampling approaches. For example, in the case of the *SSPM*, we found that the Triweight window (BMW) of size 46 exemplary projects and age 2 years resulted in the best prediction accuracy (MAE) of 0.3170 as compared to the remaining windows and the *GP* benchmark. Similarly, in the case of the *SysSam* method, we realized that the Triweight window of size 44 projects and age 1.5 years with MAE of 0.6241 constituted the *Bellwether*. Lastly, the Triweight window of size 48 projects and age 2 years with MAE of 0.8549 constituted the *Bellwether* in the case of the *RandSam* method.

We found that exemplary projects forming the *Bellwether* in the Desharnais dataset should not be older than 2 years and their window sizes should be in a range of 44 to 48 projects. In all three cases of the *Bellwether methods*, we found that the *Bellwether* yielded a relatively superior prediction accuracy as compared to the GP. Results from the Desharnais dataset thus supports the results obtained from our previous study [11], indicating that *Bellwethers* are evident and yield a relatively higher prediction accuracy against the *GP*.

2) **Kitchenham Dataset:** Results from the *SSPM, SysSam,* and *RandSam* analysis of the Kitchenham dataset are summarized in Table 6. Here, we found that among all three *Bellwether methods*, the Triweight window selected by the *SSPM* with window size of 83 exemplary projects and age of 2 years provided superior prediction accuracy (i.e., MAE of 0.5267). Thus, the Triweight window (BMW) can result in improved prediction of *new* projects as compared to the other weighted/unweighted windows and the *GP*. Results for the remaining weighted/unweighted windows and the *GP* are summarized in Table 6. With regard to the GP, we found that the Gaussian GP resulted in improved prediction accuracy as compared to the remaining weighted/unweighted GP.

As was found in our original study [11], there exists a set of exemplary projects (with defined window size and age) that yields a relatively improved prediction accuracy irrespective of the *Bellwether method* used. Also, this result validates the findings in a previous study by Kitchenham *et al.* [20] that there exists a moving window sampled from this dataset that provides better prediction accuracy as compared to the *GP*.

3) **ISBSG Dataset:** Results from the ISBSG dataset summarized in Table 7 show that the Triweight window selected by the *SSPM* yielded the best prediction accuracy as compared to those selected by *SysSam* and *RandSam*. The Triweight window (*SSPM*) has a window size of 84 exemplary projects and an age of 2 years, with a minimum MAE value of 0.6325 (see Table 7). This confirms results from our original study [11] indicating that the BMW should not be older than 2.5 years relative to the pi projects. Note that the original study used multi-organization projects from the ISBSG repository and found that the BMW should have a window size of 257 exemplary projects with an age of 2.5 years. With the exception of one Gaussian window in the case of the *SysSam* that yielded a relatively improved prediction, the Triweight window was the best across all windows.

It can be observed from Table 7 that the application of the Gaussian weighting function also resulted in improved prediction accuracy (but not as compared to Triweight) as observed in the original study [11].

Results show that there exists exemplary projects in the three studied datasets which can form the Bellwether

TABLE VII. MAE EVALUATION OF POTENTIAL BELLWETHERS SAMPLED BY *SSPM, SysSam,* AND *RandSam* AGAINST THE GROWING PORTFOLIO BENCHMARK IN THE ISBSG DATASET

| Window | | Stratification, Sampling and Prediction Method (*SSPM*) | | | | | |
|---|---|---|---|---|---|---|---|
| Age (years) | Size | Rectangular | Triangular | Epanechnikov | Gaussian | Biweight | Triweight |
| 1.5 | 82 | 1.1123 | 1.2020 | 1.4039 | 1.0244 | 1.3147 | 0.9268 |
| 2 | 83 | 1.2013 | 1.4199 | 1.8398 | 1.7749 | 1.5531 | 1.0312 |
| 2 | 84 | 1.1438 | 1.5387 | 2.0774 | 0.9234 | 1.0830 | 0.6325** |
| 2.5 | 86 | 1.0178 | 1.9018 | 1.8036 | 1.1887 | 1.0802 | 0.9829 |
| 8 | 107(GP) | 4.6061 | 3.6486 | 2.2971 | 2.8107 | 3.7094 | 2.1080 |
| Age (years) | Size | Systematic Sampling Approach (*SysSam*) | | | | | |
| 1.5 | 81 | 2.9465 | 2.6376 | 3.2752 | 2.2970 | 2.8850 | 2.4728 |
| 1.5 | 82 | 2.2105 | 2.2522 | 4.5045 | 2.8153 | 2.4635 | 2.1115 |
| 1.5 | 83 | 2.3193 | 2.3416 | 2.6832 | 1.4270 | 2.3737 | 1.3203** |
| 2 | 84 | 3.2292 | 3.0658 | 2.9053 | 2.4527 | 2.6827 | 2.2994 |
| 8 | 107(GP) | 4.6061 | 3.6486 | 2.2971 | 2.8107 | 3.7094 | 2.1080 |
| Age (years) | Size | Random Sampling Approach (*RandSam*) | | | | | |
| 1.5 | 79 | 1.6080 | 1.1351 | 2.2701 | 1.4188 | 1.2415 | 1.0641 |
| 2 | 86 | 2.0130 | 1.3918 | 2.7837 | 1.7398 | 1.5224 | 1.3049 |
| 2.5 | 86 | 3.0188 | 2.0160 | 2.0319 | 2.5200 | 2.2051 | 1.8900 |
| 2.5 | 87 | 1.7900 | 0.8453 | 1.6907 | 1.0567 | 0.9246 | 0.7925** |
| 8 | 107(GP) | 4.6061 | 3.6486 | 2.2971 | 2.8107 | 3.7094 | 2.1080 |

TABLE VIII. YUEN'S TEST AND CLIFF'S δ EFFECT SIZE FOR PAIRWISE DIFFERENCES BETWEEN THE WEIGHTED/UNWEIGHTED *Bellwether* (*SSPM*) AND *Growing Portfolio*

| Window | Trimmed mean | | Yuen's test | | Cliff's δ effect size |
|---|---|---|---|---|---|
| | LHS | RHS | t-value | p-value | |
| Desharnais dataset | | | | | |
| BMW(Tw) vs. GP(R) | 8.0903 | 15.2934 | -8.7691 | 0.0368 | 0.9243* |
| BMW(Tw) vs. GP(T) | 8.0903 | 15.7650 | -13.1157 | 0.0119 | 0.4453* |
| BMW(Tw) vs. GP(E) | 8.0903 | 24.3122 | -27.7224 | 0.0323 | 0.9925* |
| BMW(Tw) vs. GP(G) | 8.0903 | 12.4660 | 9.6117 | 0.0164 | 0.3950* |
| BMW(Tw) vs. GP(B) | 8.0903 | 16.4730 | -14.3255 | 0.0278 | 0.6120* |
| BMW(Tw) vs. GP(Tw) | 8.0903 | 14.6436 | -12.3096 | 0.0345 | 0.4324* |
| Kitchenham dataset | | | | | |
| BMW(Tw) vs. GP(R) | 7.1480 | 15.0463 | -14.3050 | 0.0178 | 0.3604* |
| BMW(Tw) vs. GP(T) | 7.1480 | 14.2752 | -17.5349 | 0.0208 | 0.9812* |
| BMW(Tw) vs. GP(E) | 7.1480 | 22.3427 | -37.3831 | 0.0362 | 0.3407* |
| BMW(Tw) vs. GP(G) | 7.1480 | 12.8766 | 8.0414 | 0.0409 | 0.4175* |
| BMW(Tw) vs. GP(B) | 7.1480 | 14.9382 | -19.1659 | 0.0362 | 0.6441* |
| BMW(Tw) vs. GP(Tw) | 7.1480 | 13.8335 | -16.4482 | 0.0273 | 0.5832* |
| ISBSG dataset | | | | | |
| BMW(Tw) vs. GP(R) | 6.4601 | 11.3314 | 9.4064 | 0.0222 | 1.2842* |
| BMW(Tw) vs. GP(T) | 6.4601 | 13.1866 | -10.5093 | 0.0149 | 0.3746* |
| BMW(Tw) vs. GP(E) | 6.4601 | 20.0357 | 8.7005 | 0.0378 | 1.4610* |
| BMW(Tw) vs. GP(G) | 6.4601 | 13.7353 | -11.4902 | 0.0118 | 0.6054* |
| BMW(Tw) vs. GP(B) | 6.4601 | 14.6537 | -7.6912 | 0.0375 | 0.4308* |
| BMW(Tw) vs. GP(Tw) | 6.4601 | 12.8210 | -10.0464 | 0.0317 | 0.6912* |

Growing portfolio (GP); Bellwether moving window (BMW); Rectangular (R); Triangular (T); Epanechnikov (E); Gaussian (G); Biweight (B); Triweight (Tw).
Statistical Significance: $p < 0.05$; Practical Significance: $* \delta \geq 0.276$.
LHS denotes the trimmed MAE of the window on the left-hand side in the first column; RHS denotes the trimmed MAE of the window on the right-hand side.

(BMW). The BMW sampled by the SSPM yielded relatively superior prediction accuracy as compared to the SysSam and RandSam methods. We found that such exemplary projects sampled by SSPM should not be older than 2 years relative to the $p_i$ hold-out projects and should have window size of 46 to 84 projects for improved prediction accuracy.

**B. Statistical Significance of Bellwethers With Respect to Weighting Functions**

We performed empirical analysis to further investigate whether *there exist statistically significant differences across the weighted/unweighted windows*. Previous studies [12], [19] had found that prediction accuracy is affected by different weighting functions. Therefore, there is a need to further investigate whether or not this assessment made in previous studies holds true in other analyses. We first used Brunner's method to test the alternative hypothesis for each of the three datasets examined. Here, we found that there exist significant differences across the weighted/unweighted windows benchmarked against the *GP*. This was evident in that the *F-values* from *Brunner's* ANOVA-like test were all significant at 5% asymptotic significance level in the three datasets: $F_1 = 332.68 (p < 0.05)$, $F_2 = 554.39 (p < 0.05)$ and $F_3 = 168.64 (p < 0.05)$ for the Desharnais, Kitchenham, and ISBSG datasets, respectively.

We further performed pairwise statistical tests between the weighted/unweighted windows and the *GP*, since Brunner's test indicated significant differences across the windows and *GP*. This was to find out if the *Bellwether* yielded improved prediction accuracy across all three datasets. The pairwise statistical tests were done using Yuen's test and Cliff's δ effect size as reported in Table 8.

Since the *Bellwether* sampled by *SSPM* yielded relatively better prediction performance (i.e., minimum MAE) on average (in Table 5), the pairwise statistical test was evaluated between *SSPM* and the *GP* (see Table 8). Results from Table 8 show that there are statistically and practically significant differences between the weighted/unweighted *Bellwether* and the *GP* across all three datasets. That is, both $p$ ($< 0.05$) and $\delta$ ($\geq 0.276$) values for Yuen's test and Cliff's $\delta$ effect size (respectively) are all significant at 5% asymptotic significance levels.

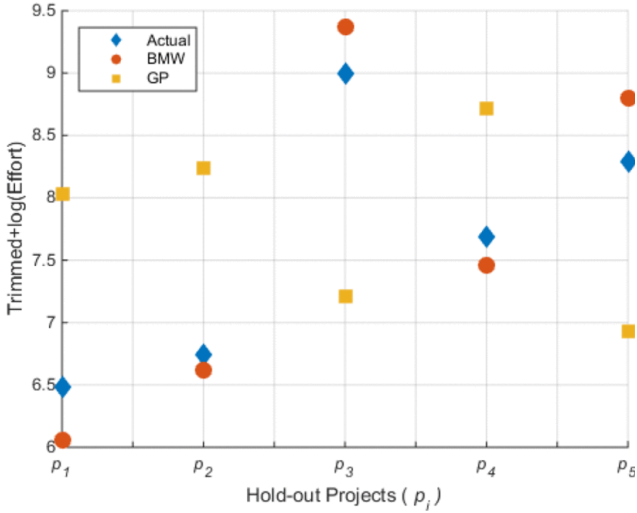

Fig. 6. Prediction of the pi hold-out projects by the *Bellwether* (BMW) and the *growing portfolio* (GP) in the Desharnais dataset. The diamond shape denotes the *actual effort* of pi as recorded in the dataset. The circular shape denotes the predicted effort of pi by the *Bellwether* (BMW) and the square shape denotes the predicted effort of pi by the *growing portfolio* (GP).

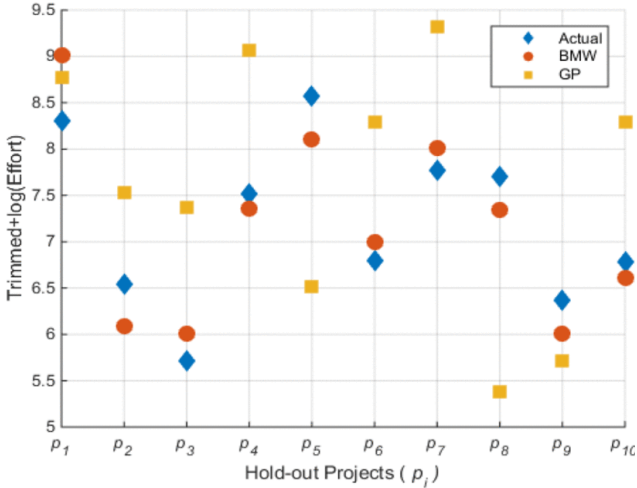

Fig. 7. Prediction of the pi hold-out projects by the *Bellwether* (BMW) and the *growing portfolio* (GP) in the Kitchenham dataset.

In order to know the *Bellwether* or *GP* that yielded the best prediction accuracy in each paired comparison test, we made use of the *trimmed means* of their respective MAE values. That is, we evaluated the prediction performances of the *Bellwether* and *GP* and averaged their respective MAE values. The minimum *trimmed mean* yielded the best prediction accuracy. Note that LHS denotes the *trimmed mean* of the weighted/unweighted *Bellwether* (BMW) and RHS denotes the *trimmed mean* of the *GP* (see Table 8). For example, in the case of the Desharnais dataset, Yuen's test and Cliff's $\delta$ effect size resulted in 0.0368 and 0.9243, respectively, significant at 5% asymptotic significance level. The *trimmed mean* for the Triweight BMW (i.e., $LHS = 8.09$) is less than that of the rectangular *GP* (i.e., $RHS = 15.29$). Therefore, the Triweight *Bellwether* yielded relatively significant performance over the rectangular *GP*. Similar results were observed for the remaining paired comparison tests.

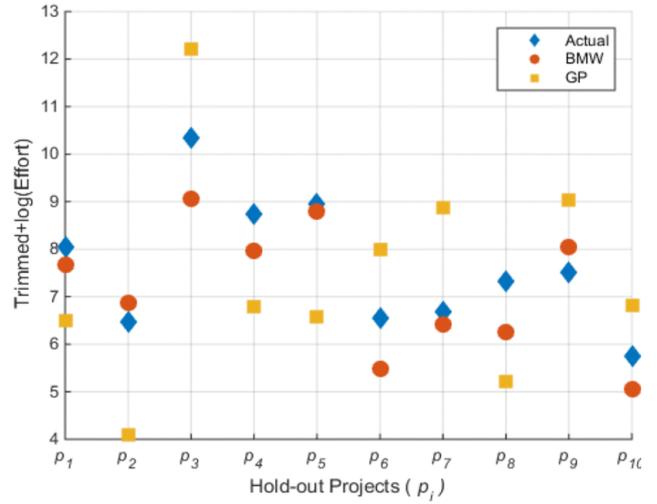

Fig. 8. Prediction of the $p_i$ hold-out projects by the *Bellwether* (BMW) and the *growing portfolio* (GP) in the ISBSG dataset.

Results show that the Triweight Bellwether (BMW) sampled by the SSPM was relatively significant wrt prediction accuracy as compared to the weighted/unweighted GP and the remaining windows. Validation of results assessed using Brunner's ANOVA-like test, Yuen's test and Cliff's δ effect size at 5% asymptotic significance level showed that Bellwethers yield superior prediction accuracy.

### C. Prediction of the $p_i$ Hold-Out Projects by the Bellwether (BMW) and the Growing Portfolio Benchmark

To investigate the significant effect of the *Bellwether* with respect to prediction accuracy, we benchmarked its prediction performance against the *GP*. This provided further evidence for considering *Bellwethers* when predicting the targets of $p_i$ projects. As recorded in Tables 5– 7, the Triweight *Bellwether* yielded the best prediction performance and hence was used for predicting the targets of the $p_i$ projects. Results are summarized in Figs. 6– 8 for the Desharnias, Kitchenham, and ISBSG datasets, respectively. Note that the *Bellwethers* sampled by the *SSPM* in each dataset were used for the prediction of the $p_i$ projects since they had resulted in superior accuracy. With regard to predicting the targets (effort) of $p_i$ in the Desharnais dataset, we found that the *Bellwether* (BMW) yielded improved prediction accuracy across all five $p_i$ projects (see Fig. 6). That is, the predicted effort by the BMW (circular shapes) resulted in similar effort values as compared to the *actual* effort values (diamond shapes) of the $p_i$ hold-out projects. Alternatively, the predicted effort of the $p_i$ projects by the GP, denoted by square shapes, resulted in less accurate predictions relative to the BMW. Similar results were observed across the Kitchenham dataset (see Fig. 7) and the ISBSG dataset (see Fig. 8).

We further performed statistical tests using Yuen's test and Cliff's δ effect size to confirm the significant differences between the prediction results of the BMW and GP relative to the actual effort values of $p_i$. As illustrated in Table 9, we found that there was no statistical and practical significance between the prediction results provided by the BMW and the actual effort values of the $p_i$ in all three datasets. For example, Yuen's *t*-test between the predicted effort by BMW and the actual effort of $p_i$ in the Desharnais dataset shows no statistically significant difference at 5% asymptotic significance level (i.e., $p = 0.9058 > 0.05$). The effect size between BMW and $p_i$ is δ = 0.1789 which is less than the threshold of 0.276, hence, the difference is not practically significant (see Table 9). Alternatively, we found statistically and practically significant differences between the GP and the $p_i$ (see Table 9). This shows that the BMW yielded superior prediction accuracy of the targets of the hold-out projects ($p_i$).

TABLE IX. STATISTICALLY SIGNIFICANT DIFFERENCES BETWEEN PREDICTION RESULTS OF THE *Bellwether* (BMW) AND THE *Growing Portfolio* (GP) RELATIVE TO $p_i$

| BMW/GP vs $p_i$ | Yuen's test | | Cliff's δ effect size |
|---|---|---|---|
| | *t-test* | *p-value* | |
| **Desharnais dataset** | | | |
| BMW vs $p_i$ | 0.1260 | 0.9058 | 0.1789 |
| GP vs $p_i$ | 3.4149 | 0.0269* | 0.8748** |
| BMW vs GP | -2.8552 | 0.0462* | 1.0384** |
| **Kitchenham dataset** | | | |
| BMW vs $p_i$ | -0.4416 | 0.6692 | 0.1241 |
| GP vs $p_i$ | 2.8966 | 0.0177* | 0.4898** |
| BMW vs GP | -3.3441 | 0.0086* | 0.4407** |
| **ISBSG dataset** | | | |
| BMW vs $p_i$ | -1.3263 | 0.2174 | 0.2113 |
| GP vs $p_i$ | 2.8355 | 0.0195* | 0.6265** |
| BMW vs GP | -3.0039 | 0.0149* | 0.6847** |

## 8. GUIDELINE FOR SAMPLING BELLWETHER MOVING WINDOWS

In this section, we introduce the following guideline for sampling BMW to be considered for predictive modeling. These guidelines are validated using the three studied datasets.

1. The BMW will continue to be useful for prediction purposes provided the phenomenon being measured shares similar characteristics with the exemplary projects. In other words, if the *new* project to be developed does not share similar characteristics (such

2. as the number of features, development, and testing procedures) with the projects within the BMW, then the BMW might not be useful for predictive modeling.

3. The feature set, $f_{new}$ of the *new* project whose software effort is to be predicted should be a subset of the feature set, $f_{bell}$ of the BMW. That is, $f_{new} \subseteq f_{bell}$. Thus, $f_{new}$ should have the same or a subset of *prior features* in $f_{bell}$ when seeking to use the BMW to predict the software effort of *new* project(s). For example, given that the *prior features* such as $f_{bell}^{(1)}, \ldots, f_{bell}^{(n)}$ are considered during the training/validation process for obtaining the BMW, then the features, $f_{new}$ of the *new* project should have the same features or a subset of features considered in the selection of the BMW. That is, $f_{new}^{(i)} \subseteq f_{bell}^{(j)}$ where $i \leq j$ and $i, j \neq k$. $k$ is a set of *posterior* features (e.g., LOC, number of source files and the like) that is not considered during the training/validation process, and therefore cannot be used when predicting the software effort of *new* project(s).

4. *Posterior* features are not considered for sampling the BMW since the *new* project to be predicted does not share such characteristics in the initial stage of development. We therefore recommend the use of *prior* features for obtaining the BMW. *Prior* features refer to the independent variables known before (or at the beginning of) the development of a *new* project, whilst *posterior* features refer to the independent variables known after the development of a project [25]. Examples of *prior* features are function points, number of developers, experience of development team, development platform, etc., and examples of *posterior* features are lines of code (LOC), number of source files, total number of commits, etc.

5. Historical project datasets can be *noisy* and heterogeneous which affects data quality [81] and as such, there is a need to preprocess the data [59] prior to sampling the BMW. We recommend the use of *trimming + log transformation* as an effective data preprocessing method for obtaining the BMW.

6. Historical project datasets collected from different industries or organizations are heterogeneous in nature with different methodological, development, and security policies. As a result, single organizational projects which share similar characteristics with the $p_i$ hold-out projects are to be preferred for sampling the BMW for improved prediction accuracy. In cases where an organization lacks historical projects, the introduced *Bellwether methods* can be adopted to sample cross-company projects as shown using the Desharnais dataset (in this study) and the ISBSG dataset (in the original study [11]).

7. From our empirical investigation of the existence of *Bellwethers* in chronological datasets, we found that such exemplary projects exist in at least 71 project cases (*GP* size of the preprocessed Desharnais dataset).

8. Results from this study have shown that the window age of the BMW should not be older than 2 years. Even though we still have limited empirical evidence based on sampling from different chronological datasets, we recommend that in the use of the studied datasets, the age of the BMW should not exceed 2 years.

9. With regard to window size of the BMW, we found an approximate range of 50 to 80 exemplary projects. This range has proven to form a relatively reasonable window size for predicting the target of *new* project data based on our empirical evidence.

## 9. RELATED WORK

To the best of our knowledge, Kitchenham *et al.* [20] were the first to consider the selection of recently completed projects (*moving windows*) from a chronologically ordered project dataset for SEP. They made use of the (subsequently named) Kitchenham dataset comprising projects obtained from a single organization. They divided the *GP* into four partitions and each partition was used as the training set for modeling. Results from their regression analysis show that the size to effort estimation relationship changed across the four partitions. They concluded that older projects should be removed from the dataset and prediction done with the remaining subset together with the addition of recently completed projects. Lastly, they recommended the prediction of *new* projects with respect to the use of about 30 recently completed projects.

A recent study by Amasaki and Lokan [12] investigated the effect of weighted moving windows on SEP accuracy using the Finnish dataset. They found that the application of different weighting functions has different effects on the SEP accuracy. They concluded that weighted moving windows with comparably larger window size yielded significantly better accuracy as compared to unweighted moving windows. Their findings confirmed results from their previous study [15] where similar outcomes were found using the ISBSG dataset. Amasaki and Lokan [12] considered window sizes ranging between 20 and 120 projects (i.e., 20, 30, 40, …, 120) in building their prediction models based on weighted linear regression. Results show that models built with windows of sizes ranging from 20 to 30 projects yielded minimum MAE as compared to the use of the *GP* in SEP modeling. The prediction accuracy diminished when using triangular, epanechnikov. and Gaussian weighting functions for window sizes below 80 projects whilst the unweighted moving window (Rectangular) resulted in better accuracy. Lastly, with window sizes more than 80 projects, prediction accuracy was highly improved when using the three weighted moving windows (triangular, epanechnikov, and Gaussian) as compared to using unweighted moving windows. Results from their study [12] showed that the use of moving windows from the Finnish dataset could improve prediction accuracy, a contradiction to a recent study by Lokan and Mendes [14]. They [14] reported that companies should not discard older projects but rather continue to use them for the training of prediction models instead of using moving windows.

A further study by Amasaki and Lokan [15] examined the effects of weighted moving windows and unweighted moving windows on prediction accuracy. They also considered projects from the ISBSG dataset (release 10) and used weighted linear regression in building the SEP model. Results from their study show that

1. unweighted moving windows are more effective in SEP with larger window size (at least 40 projects) as compared to an unweighted *GP*;

2. a weighted *GP* of large size yielded better prediction accuracy as compared to an unweighted *GP,* with respect to MAE;

3. an unweighted moving window is more effective in prediction accuracy than a weighted *GP*;

4. for large window size of at least 100 projects, a weighted moving window yielded better prediction accuracy than an unweighted *GP* for both MAE and mean magnitude of relative error (MMRE). Here, the Gaussian weighting function was considered the best;

5. weighted moving windows of large sizes are better than weighted *GPs* with respect to MAE and MMRE. The Gaussian function was considered the best weighting function in this case;

6. weighted moving windows with large sizes (at least 80 projects) are more effective in terms of accuracy than unweighted moving windows for MAE and MMRE. Here, no preference to the best weighting function was made in their study. Amasaki and Lokan [15] concluded that using windows of large size (more than 75 projects) improves prediction accuracy.

Lokan and Mendes [13] investigated the effect of using moving windows based on different durations and how it affects prediction accuracy. The ISBSG dataset (release 10) and the Finnish dataset were considered in their study. They found that prediction accuracy can be improved when using moving windows based on duration. Here, stepwise multivariate regression was used in building the SEP model and windows of different durations ranging from 1 to 7 years were considered in modeling. Results show that SEP models built with moving windows of about 3 years with at least 75 projects (in the training set) yielded superior prediction accuracy. Similar results were found in their previous study [17] whereby moving windows of about 3 to 4 years yielded improved prediction accuracy with a relative effective size of 81 to 89 projects using the ISBSG dataset.

This study acknowledges the assumption made by previous studies that moving windows can improve prediction accuracy. However, since prediction accuracy is affected by different window sizes and ages, there is the need for further investigation to address the aging and sizing constraint. This paper introduced three *Bellwether methods* (*SSPM, SysSam,* and *RandSam*) to select exemplary projects based on the concept of the *Bellwether effect* and Markov chains to address the sizing and aging constraints of moving windows. These exemplary projects referred to as *Bellwethers* are obtained using well-defined stratification and sampling procedures (see Section 5). The *Bellwethers* can be considered as the best training set for constructing an SEP model to predict the targets of *new* projects.

## 10. THREATS TO VALIDITY

### A. External Validity

We considered historical projects from three chronological datasets. These datasets are convenience samples and so cannot be considered as generally representative of all chronological datasets. Therefore, results from this study might not be confidently generalized beyond these datasets. However, the datasets used in this study are relatively large and consist of projects from well-known data repositories (i.e., ISBSG and PROMISE [21]). Hence, they can be considered as a fair and generalizable representation of projects for any organization that develops projects bearing similar characteristics.

As a result of using chronological projects for this study, it is important to note that numerous projects without start and completion dates were eliminated during the preprocessing phase. Thus, the heterogeneous nature of these datasets especially the ISBSG dataset resulted in a vast reduction of sampled projects. Hence, this cannot lead to a true reflection of results provided all projects had known start and completion dates.

**B. Construct Validity**

We considered a single learner, namely DNN [25] for building the prediction model. We chose this learner because it had been shown to improve prediction accuracy in our original study [11] and hence could be used as a reliable and effective SEP model. Even though this study is limited to a single learner for the training and validation needs, the DNN model might yield similar prediction results to other SEP models used in previous studies [12], [14], [15].

In this study, we considered a single evaluation measure, namely MAE. Even though other evaluation measures could have been used for evaluation assessment, the MAE has been proven reliable and effective in previous studies [3], [24], [33], [36]. MAE is an unbiased performance measure towards model overestimation and underestimation [36], [14], [33]. Other evaluation measures such as the MMRE, median magnitude of relative error, and PRED($k$) are misleading evaluation measures which can result in bias in model assessment [36].

**C. Conclusion Validity**

The models employed in this study were automated and hence there is a risk of automation bias. For example, the sampling of *Bellwethers* by *SSPM*, *SysSam,* and *RandSam* was done automatically. Automating a process involves a series of assumptions being made; nevertheless, in this case those assumptions are based on prior knowledge from previous studies [10], [12], [14], [15], [40], [82] in building data mining and prediction models. Hence, there is some basis for believing that the results can be trusted.

In summary, robust statistical tests (e.g., Yuen's test and Brunner's test) suitable for the datasets and problem at hand were carefully applied to assure the conclusion validity of results. In addition, a robust effect size method, namely Cliff's δ effect size, was used to provide additional evidence as to the significance and relevance of our results.

## 11. Summary, Conclusion, and Future Works

Selection of relevant training data from a pool of historical and chronological datasets to be considered as a *moving window* has been investigated in previous studies [12], [14], [15], [18] with the goal of improving SEP accuracy. The use of exemplary and recently completed projects (*Bellwether moving window*) was found in a previous study [11] to further improve on prediction accuracy in SEP. However, the different sizing, aging and weighting function parameters of the *Bellwether moving window* themselves affect prediction accuracy. This study sought to tackle the three aforementioned constraints by conducting empirical and theoretical investigations using three *Bellwether methods* (*SSPM, SysSam* and *RandSam*) to sample *Bellwethers* from three chronological datasets. The selected *Bellwethers* were benchmarked against the *GP* (the entire collection of historical projects) and were used to predict the targets of $p_i$ hold-out projects. The study also introduced two weighting functions (Biweight and Triweight functions) in addition to the existing four weighting functions [15] to further investigate their effects on prediction accuracy. The sizing and aging constraints of the *Bellwether* were addressed based on the Markov chain Monte Carlo (MCMC) methodology. MCMC is based on a probabilistic principle that the prediction of a future event is independent of past events given the present event. Investigations of the three studied chronological datasets (Desharnais, Kitchenham and ISBSG) reinforce the existence of exemplary projects (referred to as the *Bellwether effect*) to be used for the training needs of prediction models. Validation of the study's results was performed using robust statistical tests, namely Brunner's test, Yuen's test and Cliff's δ effect size, as recommended by Kitchenham *et al.* [31]. Statistical inferences were made at the 5% asymptotic significance level.

We also contributed with a guideline (based on the studied datasets) to assist researchers and practitioners in sampling *Bellwethers* to be used in training prediction models. Results from this study have shown that the *Bellwether moving window* should have an approximate window size of 50 to 80 projects and should not be older than 2 years old relative to the *new* projects. We observed that weighting the *Bellwether moving window* with the Triweight function was more advantageous with respect to relative prediction accuracy. The characteristics of the exemplary projects that constituted the *Bellwether moving window* were further investigated and we found that *trimming + log transforming* the data yielded improved prediction accuracy as compared to using untransformed data.

In a future study, we seek to conduct empirical studies in other software engineering domains (such as software defect prediction) to further investigate if the *Bellwether moving window* will yield improved prediction accuracy over the *GP*. A further study on this research is to compare results with other models such as the Bayesian estimation model and other deep learning models (recurrent and convolutional neural networks). we intend to extend the *Bellwether* concept to other SEP chronological datasets such as the Finnish dataset, Maxwell dataset, and other industrial datasets. This will add more weight in generalizing the relative effectiveness of using *Bellwethers* for improved prediction accuracy in software engineering and other related fields.

**Solomon Mensah** (S'17) received the B.Sc. degree in computer science and statistics and the M.Eng. degree in computer engineering from the University of Ghana, Accra, Ghana, in 2011 and 2014, respectively. He is currently working toward the Ph.D. degree in computer science at the Department of Computer Science, City University of Hong Kong, under the supervision of Dr. Jacky Keung.

His research interests include software effort prediction, statistical modeling, deep learning, and technical debt.

**Jacky Keung** (M'18) received the B.Sc. (Hons.) degree in computer science from the University of Sydney, Australia, and the Ph.D. degree in software en- gineering from the University of New South Wales, Australia.

He is an Assistant Professor with the Department of Computer Science, City University of Hong Kong, Hong Kong. His main research area is in software effort and cost estimation, empirical modelling and evaluation of complex sys- tems, and intensive data mining for software engineering datasets. His research has been published in prestigious journals including IEEE Transactions on Software Engineering, Empirical Software Engineering Journal, Information and Software Technology, the Journal of Systems and Software, Automated Software Engineering and many other leading journals and conferences.

Dr. Keung is also the vice-president of IEEE Hong Kong Section Computer Society Chapter.

**Stephen G. MacDonell** (SM'17) received the B.Com. (Hons.) degree in in- formation systems from the University of Otago, Dunedin, New Zealand, in 1988, the M.Com. degree in information science from the University of Otago, Dunedin, New Zealand, in 1990, and the Ph.D. degree in engineering from the University of Cambridge, Cambridge. U.K., in 1993.

He is a Professor of Software Engineering with the Auckland University of Technology, Auckland, New Zealand, and a Professor in Information Science with the University of Otago. His research has been published in the IEEE TRANSACTIONS ON SOFTWARE ENGINEERING, *ACM Transactions on Software Engineering and Methodology*, *ACM Computing Surveys*, *Empirical Software Engineering*, *Information & Management*, *Information and Software Technol- ogy*, the *Journal of Systems and Software*, and the *Project Management Journal*, and he has presented his research findings at numerous international confer- ences.

Prof. MacDonell is a Fellow of IT Professionals NZ, a Senior Member of the IEEE Computer Society, and a Member of the ACM, and he serves on the Editorial Board of *Information and Software Technology*. He is also Theme Leader for IT, Data Analytics and Modelling in New Zealand's National Science Challenge *Science for Technological Innovation*.

**Michael Franklin Bosu** received the B.Sc. (Hons) in computer science from the Kwame Nkrumah University of Science and Technology, Ghana in the year 2000. He was awarded M.Sc. Computer Engineering and M.Sc. Engineering and Management of Information Systems by Dalarna University, Sweden in 2003 and the Royal Institute of Technology, Sweden in 2006 respectively and a Ph.D. in Information Science from the University of Otago, New Zealand in 2016.

He is currently a Lecturer with the Waikato Institute of Technology, Hamil- ton, New Zealand, and leads the Database Subject Group. His main research interests include data quality in empirical software engineering, software effort estimation, and mining of application reviews. He has published in reputable journals and presented at conferences.

**Kwabena Ebo Bennin** (S'16) received the B.Sc. (Hons.) degree in computer science and statistics from the University of Ghana, Accra, Ghana, in 2011. He is currently working toward the Ph.D. degree in computer science at the Department of Computer Science, City University of Hong Kong. He is under the supervision of Dr. Jacky Keung.

He was a visiting researcher with Okayama University, Japan, working under the supervision of Prof. Akito Monden. His research interests include improving the fundamentals and performances of software defect prediction models, soft- ware effort and cost estimation, data mining for software engineering datasets, and techniques for bug localization. Mr. Bennin is an HKPFS fellow.